\documentclass[10pt,letterpaper]{article}
\usepackage[top=0.85in,left=1.75in,footskip=0.75in,marginparwidth=2in]{geometry}

\usepackage[utf8]{inputenc}

\usepackage{cite}
\usepackage{amsthm, amsmath, amssymb}
\usepackage{graphicx,subcaption}
\usepackage{changepage}
\usepackage[title]{appendix}
\usepackage{float}
\usepackage{tabularx}
\usepackage{booktabs}
\usepackage{nameref,hyperref}

\usepackage[right]{lineno}

\usepackage{microtype}
\DisableLigatures[f]{encoding = *, family = * }

\usepackage{changepage}

\usepackage[aboveskip=1pt,labelfont=bf,labelsep=period,singlelinecheck=off]{caption}

\usepackage{comment}
\makeatletter
\renewcommand{\@biblabel}[1]{\quad#1.}
\makeatother

\usepackage{lastpage,fancyhdr,graphicx}
\usepackage{epstopdf}
\fancyheadoffset[L]{2.25in}
\fancyfootoffset[L]{2.25in}

\usepackage{color}

\definecolor{Gray}{gray}{.25}

\usepackage{graphicx}

\usepackage{sidecap}

\usepackage{wrapfig}
\usepackage[pscoord]{eso-pic}
\usepackage[fulladjust]{marginnote}
\reversemarginpar

\begin{document}
\vspace*{0.35in}

\begin{flushleft}
{\Large
\textbf\newline{Multistable protocells can aid the evolution of prebiotic autocatalytic sets}
}
\newline
\\
Angad Yuvraj Singh \textsuperscript{1},
Sanjay Jain\textsuperscript{1,2*}
\\
\bigskip
\bf{1} Department of Physics and Astrophysics, University of Delhi, Delhi 110007 India
\\
\bf{2} Santa Fe Institute, 1399 Hyde Park Road, Santa Fe, New Mexico 87501, USA
\\
\bigskip
* jain@physics.du.ac.in

\end{flushleft}

\section*{Abstract}
We present a simple mathematical model that captures the evolutionary capabilities of a prebiotic compartment or protocell. In the model the protocell contains an autocatalytic set whose chemical dynamics is coupled to the growth-division dynamics of the compartment. Bistability in the dynamics of the autocatalytic set results in a protocell that can exist with two distinct growth rates. Stochasticity in chemical reactions plays the role of mutations and causes transitions from one growth regime to another. 
We show that the system exhibits `natural selection', where a `mutant' protocell in which the autocatalytic set is active arises by chance in a population of inactive protocells, and then takes over the population because of its higher growth rate or `fitness'. The work integrates three levels of dynamics: intracellular chemical, single protocell, and population (or ecosystem) of protocells..


\section*{Introduction}
The simplest life forms existing today and plausibly existing at the origin of life are such complex chemical organizations involving small and large molecules, that it is virtually impossible to imagine their origin except through some process of chemical evolution \cite{oparin1924,haldane1929}. Imagining plausible steps in chemical evolution that resulted in the increase of complexity of prebiotic chemical organization is therefore an important task. 

One significant set of prebiotic scenarios is based on the idea of an autocatalytic set (ACS) of chemical reactions \cite{eigen,kauffman1971,rossler}, reviewed in \cite{origins,hordijkreview}. Here we are concerned about the evolution of ACSs. This has been investigated \cite{bagleyfarmer2,jainkrishna1998, vasas2012} (for reviews, see \cite{nghe2015,ametareview}) largely in the context of ACSs that reside in static well stirred containers. It is recognized that at some stage autocatalytic networks must have evolved inside a spatial compartment or `protocell' which propagated through growth and division. Consequently, different models of protocells containing ACSs have been proposed \cite{ganti1975, gard1998, rasmussen2003, sole2007, mavelli2007, carletti2008, kaneko2010, hordijk2018, luisi_book, serra_book} where the compartments are modeled after micelles (autocatalytic aggregates of lipid catalysts), vesicles (lipid bilayers permeable only to food molecules enclosing an aqueous environment containing the ACS) or other structures.  

These models have considered how the features of Darwinian evolution \cite{lewontin1970,godfrey2007}, namely, (i) heredity, (ii) heritable variation, and (iii) differential fitness of the variants, can arise in such protocells. 
In models of growing-dividing protocells that contain ACSs, daughter protocells inherit the composition of the mother, and this transmission of compositional information is the mechanism of heredity \cite{gard2000,vasas2012, villani2014} instead of template replication of an information carrying molecule. The interesting property of `synchronization' has been shown to arise fairly generically in these models \cite{carletti2008, serra2019} whereby the composition of the protocell at successive divisions remains the same, giving the lineage of protocells a stable compositional identity. As a source of variation needed for evolution, models have considered chemical fluctuations due to the chance occurrence of rare reactions which are enhanced in small volumes, or changes in the environment (e.g., addition or removal of molecular species from the food set) \cite{togashi2001,serra2014,hordijk2018,kahana2023}. 
A large network containing multiple ACSs \cite{hordijk2004,blokhuis2020} 
causes protocells that contain distinct ACSs to grow with different rates \cite{vasas2012,villani2014}. This
can give rise to differential fitness of protocells. 

Notwithstanding all the above work, a crisp and convincing theoretical demonstration of the Darwinian evolution of a population of ACS containing protocells remains an unfinished task \cite{ametareview}. In this paper we present a new model which explicitly demonstrates the evolution of a population of such protocells in the Darwinian sense (albeit only one step of evolution due to the simplicity of the model). Our work makes use of 
an interesting feature of certain autocatalytic network topologies: the presence of multi-stability in the dynamics \cite{ohtsuki2009,wu2009,piedrafita2010, giri2012,matsubara2016}. 
Our protocell has just two stable states, one in which no ACS is present (inactive state) and the other in which it is (active state). The protocell has a higher growth rate in the active state compared to the inactive state. The variation in a protocell is just the spontaneous transition, due to chemical fluctuation in a small volume, from the inactive to the active state without any change of environment. The evolution exhibited is the establishment, growth and dominance of the active protocells in a population of protocells. The simplicity of the model allows us to quantify the conditions under which this `natural selection' can take place, in terms of the various dynamically generated timescales of the model. In future work we hope to generalize this to multiple evolutionary steps of increasing complexity.

\section{The model}
The protocell consists of three molecular species, a monomer $A(1)$ (food molecule), a dimer $A(2)$ (assumed to be the enclosure forming molecule) and tetramer $A(4)$ (catalyst); see Fig. \ref{fig1}. The population of $A(i)$ $(i=1,2,4)$ in the protocell is denoted $X_i$; $x_i \equiv X_i/V$ is its concentration, where $V$ is the volume of the protocell. The set of reactions these molecules can undergo are:

\begin{figure}[t]
\centering
\includegraphics[width=10cm]{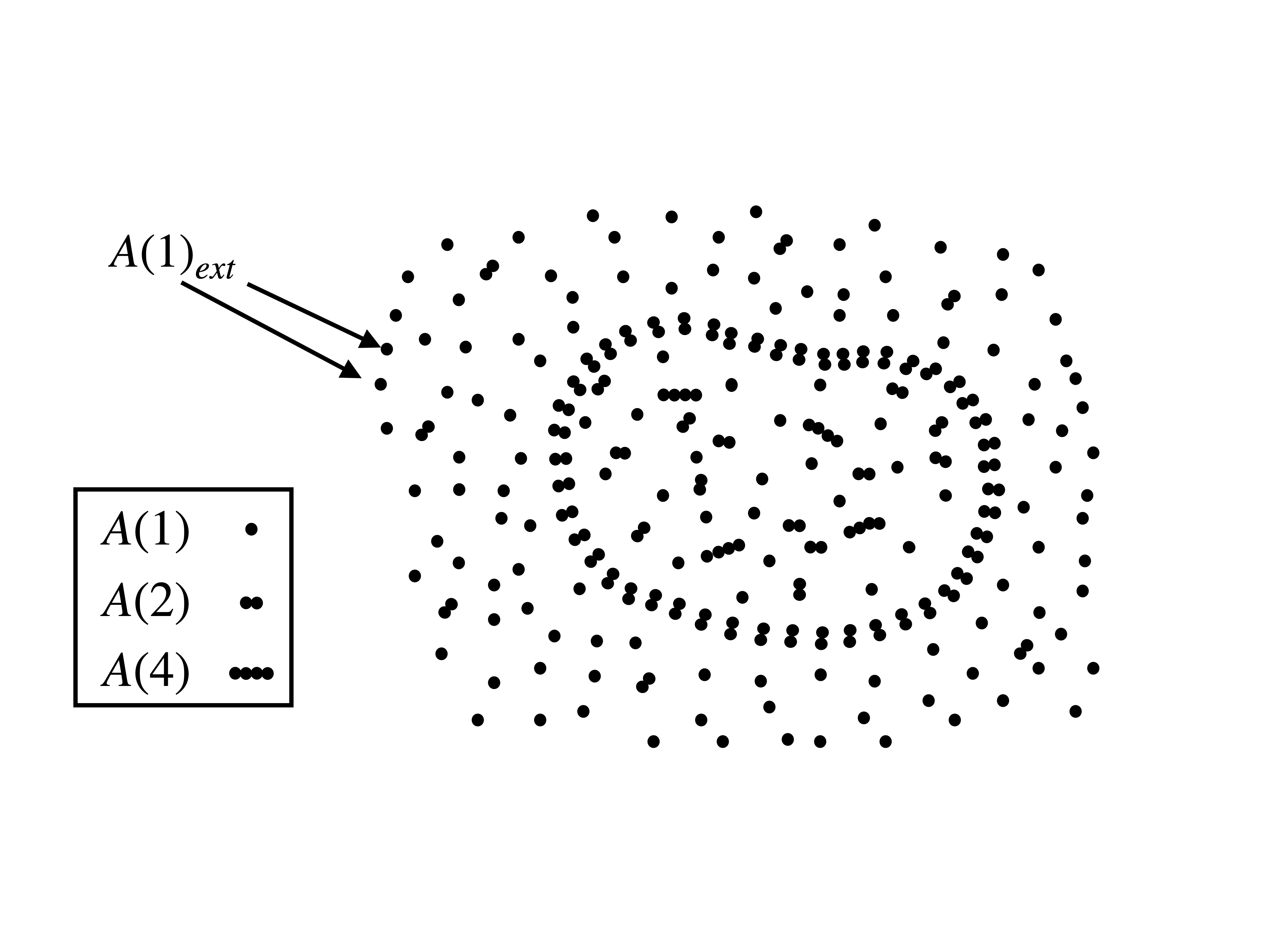}
\caption{\small{An illustration of a protocell inside an aqueous medium buffered with monomeric food molecules, $A(1)_{ext}$. The protocell membrane is composed of dimer molecules $A(2)$. }}
\label{fig1}
\end{figure}

\begin{align*}
&\text{\bf Transport}:\;&A(1)_{ext}&\overset{\alpha X_2}\longrightarrow A(1)\\
&\text{\bf R1 (uncatalyzed)}:\;&2A(1)\;&\overset{k_F}{\underset{k_R}{\rightleftharpoons}} \;\;A(2)\\
&\text{\bf R1 (catalyzed)}:\;&2A(1)+ A(4)\; &\overset{\kappa k_F}{\underset{\kappa k_R}{\rightleftharpoons}} \;\;A(2)+A(4)\\
&\text{\bf R2 (uncatalyzed)}:\;&2A(2)\;&\overset{k_F}{\underset{k_R}{\rightleftharpoons}} \;\;A(4)\\
&\text{\bf R2 (catalyzed)}:\;&2A(2)+ A(4)\; &\overset{\kappa k_F}{\underset{\kappa k_R}{\rightleftharpoons}} \;\;A(4)+A(4)\\
&\text{\bf Degradation}:\;&A(2)\overset{\phi}\longrightarrow \emptyset,&\;\;A(4)\overset{\phi}\longrightarrow \emptyset.
\end{align*}

\noindent $A(1)_{ext}$ denotes the monomer species outside the cell; its concentration is assumed constant. The membrane formed by the dimers is permeable only to monomers; the rate at which monomers come in is proportional to the number of dimers, $\alpha$ being the proportionality constant. Two monomers can spontaneously ligate to form a dimer and two dimers to form a tetramer, both with the same rate constant $k_F$. The reverse (dissociation) reactions have a spontaneous rate constant $k_R$. These ligation-dissociation reactions are also catalyzed by the tetramer, whose `catalytic efficiency' is denoted $\kappa$ (this effectively means that the catalyzed reaction rate is $\kappa x_4$ times the spontaneous rate). The dimer and tetramer are assumed to degrade with rate constant $\phi$ into a waste product that quickly diffuses out of the protocell. Note that the catalyzed reactions R1 and R2 together with the transport reaction form an ACS starting from the food set $A(1)_{ext}$. 

In this model the dimer does double duty as both the enclosure forming molecule as well as a reactant for catalyst production. In the equations below, we do not introduce separate population variables for the two roles. This is purely for simplicity and is not a crucial assumption. In the Supplementary Material Section 1 we show that in a model with two monomer species in which these two functions are performed by distinct molecules, similar results arise. 

Using mass action kinetics, the deterministic rate equations of the model are given by
\begin{align} 
\frac{dx_1}{dt}=&\;\alpha x_2 - \:2(k_F' x_1^{2}-k_R' x_{2})- \frac{\dot V}{V} x_1,\label{eq_conc1} \\ 
\frac{dx_2}{dt}=&\;k_F' x_1^{2}-k_R' x_{2}\: \nonumber \\&- \:2(k_F' x_2^2-k_R'x_{4}) \:-\:(\phi+\frac{\dot V}{V}) x_2, \\
\frac{dx_4}{dt}=&\;(k_F'x_2^{ 2}-k_R'x_{4}) \:-\:(\phi+\frac{\dot V}{V}) x_4,\label{eq_conc3} \\
k_F' \equiv &\;k_F(1 + \kappa x_4), \; k_R' \equiv k_R(1+ \kappa x_4).
\end{align}

\noindent The $\dot{V}/V$ terms represent dilution in an expanding volume. Note that when $V$ is not constant, Eqs. (\ref{eq_conc1}-\ref{eq_conc3}) do not specify the dynamics 
completely unless the growth rate $\dot{V}/V$ is specified. Since here we want an endogenous growth rate, we do not specify $\dot{V}/V$ exogenously. Instead, we write the model in terms of the populations, and assume a certain functional form for $V$ in terms of the populations. In terms of $X_i$, the above equations reduce to
 \begin{align}
\frac{dX_1}{dt}=&\alpha X_2 - \:2(\frac{k_FX_1^2}{V}-k_R X_{2})(1+\kappa \frac{X_4}{V}),\label{eq_pop1} \\ 
\frac{dX_2}{dt}=&(\frac{k_F X_1^2}{V}-k_RX_{2})(1+\kappa \frac{X_4}{V})\nonumber \\
&\:- \:2(\frac{k_F X_2^2}{V}-k_RX_{4})(1+\kappa \frac{X_4}{V}) \:-\:\phi X_2,\label{eq_pop2} \\
\frac{dX_4}{dt}=&(\frac{k_F X_2^2}{V}-k_RX_{4})(1+\kappa \frac{X_4}{V}) \:-\:\phi X_4. \label{eq_pop3}
\end{align}
For simplicity we take $V$ to be a linear function of the populations $X=(X_1,X_2,X_4)$:
\begin{equation}
V(X) = v(X_1+2X_2+4X_4), \label{eq_vol}
\end{equation}
where $v$ is a constant. This choice gives the protocell a constant mass density (as observed in bacterial cells \cite{martinez-salas1981}) since $V$ is proportional to the mass of the protocell. This choice is not essential; we have tried other linear functions $V = v_1X_1 + v_2X_2 +v_4X_4$ ($v_i$ constant), including $V=v(X_1 + X_2 + X_4)$. The quantitative results depend on the values of $v_i$ but the qualitative features presented below hold for all the cases considered. We have also considered other versions of the model with the transport term $\alpha X_2$ in \ref{eq_pop1} modified to a gradient term $\alpha X_2(x_{1, \text{ext}} - x_1)$ (where $x_{1, \text{ext}}$ is the constant concentration of $A(1)_{ext}$), certain other autocatalytic reaction topologies, etc. (see Supplementary Material Section 1). The qualitative conclusions seem to be robust to these choices. Without loss of generality, the constants $k_R$ and $v$ are set to unity by rescaling $t \rightarrow k_Rt$, $\alpha \rightarrow  \alpha/k_R$, $\phi \rightarrow \phi/k_R$, $k_F \rightarrow k_F/(k_Rv)$, $\kappa \rightarrow \kappa/v$, which makes time and the other parameters dimensionless. 

The definition of $V(X)$ and the values of the rescaled parameters $k_F, \phi, \alpha, \kappa$ completely define Eqs. (\ref{eq_pop1}-\ref{eq_pop3}), and one can solve for $X(t)$ given any initial condition. In a particular trajectory $V$ may increase or decrease. Protocells larger than a characteristic size may become floppy or unstable and spontaneously break up into smaller entities. We assume that if $V$ increases to a critical value $V_c$ the cell divides into two identical daughters each containing half of the three chemicals of the mother protocell at division. The dynamics of a daughter after division is again governed by Eqs. (\ref{eq_pop1}-\ref{eq_vol}). This division rule and Eqs. (\ref{eq_pop1}-\ref{eq_vol}) together completely define the model at the deterministic level.

The dynamics of the ACS consisting of the catalyzed reactions R1 and R2 in a fixed size container but with buffered $A(1)$ as the food set is given by Eqs. (\ref{eq_pop2}-\ref{eq_pop3}) with $V$ and $X_1$ constant. This was studied in \cite{giri2012} at the deterministic level where a bistability was observed, and in \cite{matsubara2016} at the stochastic level where transitions between the attractors was observed. The present model by adding Eqs. (\ref{eq_pop1}), (\ref{eq_vol}) and the division rule embeds the ACS in a growing-dividing protocell instead of a fixed volume container. It shares the bistability of the fixed volume version, but also possesses qualitatively new properties. These properties (considered along with stochastic dynamics) enable a population of such protocells to mimic (one step of) Darwinian evolution, as will be discussed below.

\section{Results}
\subsection{Deterministic dynamics: Bistability with two distinct growth rates}

Since $V$ is a linear function of the populations, $\dot{V}/V$ can be expressed in terms of the concentrations. Differentiating Eq. (\ref{eq_vol}) w.r.t. $t$ and using Eqs. (\ref{eq_pop1}-\ref{eq_pop3}), it follows that 
\begin{align}
\mu \equiv \frac{\dot{V}}{V}=\alpha x_2-\phi(2x_2 + 4x_4). \label{eq_mu}
\end{align}
Eqn. (\ref{eq_mu}) expresses the instantaneous growth rate of the protocell in terms of its chemical composition, a feature that is missing from previous protocell models.

\begin{figure}[t]
\centering
\includegraphics[width=\textwidth]{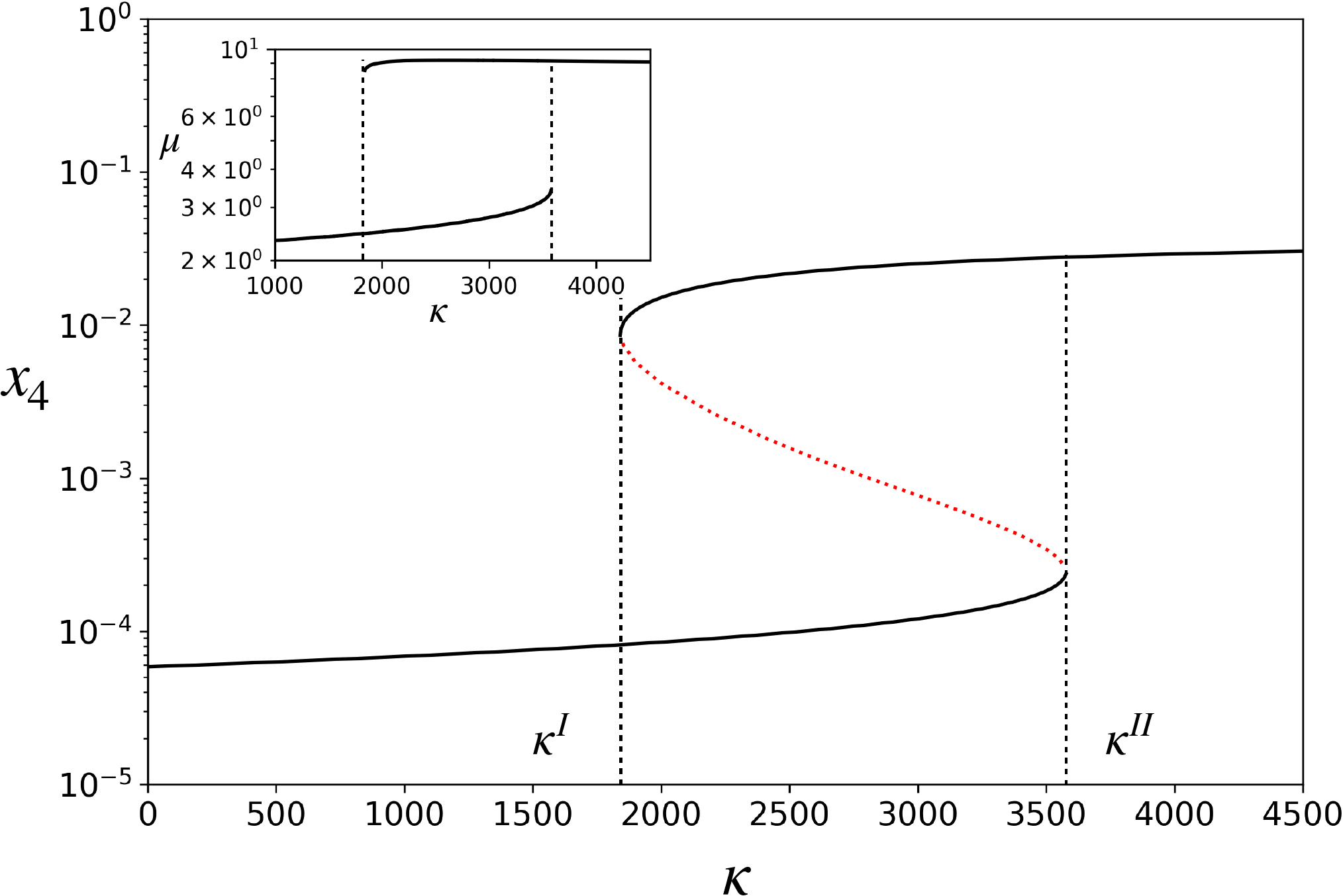}
\caption{\small{Bifurcation diagram for the model: Steady state concentration, $x_4$, of the catalyst versus catalytic efficiency, $\kappa$. The region between $\kappa^I (= 1840)$ and $\kappa^{II} (=3580)$ is the region having three fixed points, two of which are stable (solid black curves) and one is unstable (red dotted curve). \textbf{Inset}: Growth rate, $\mu$, of the protocell, versus  $\kappa$. Parameters: Hereafter, $k_R$ and $v$ have been set to unity without loss of generality after non-dimensionalizing the model. $k_F = 1$, $\phi=20$, $\alpha=100$. }}
\label{fig2}
\end{figure}

\begin{figure}[t]
\begin{adjustwidth}{-1.5cm}{-1.5cm}
\centering
\includegraphics[width=18cm]{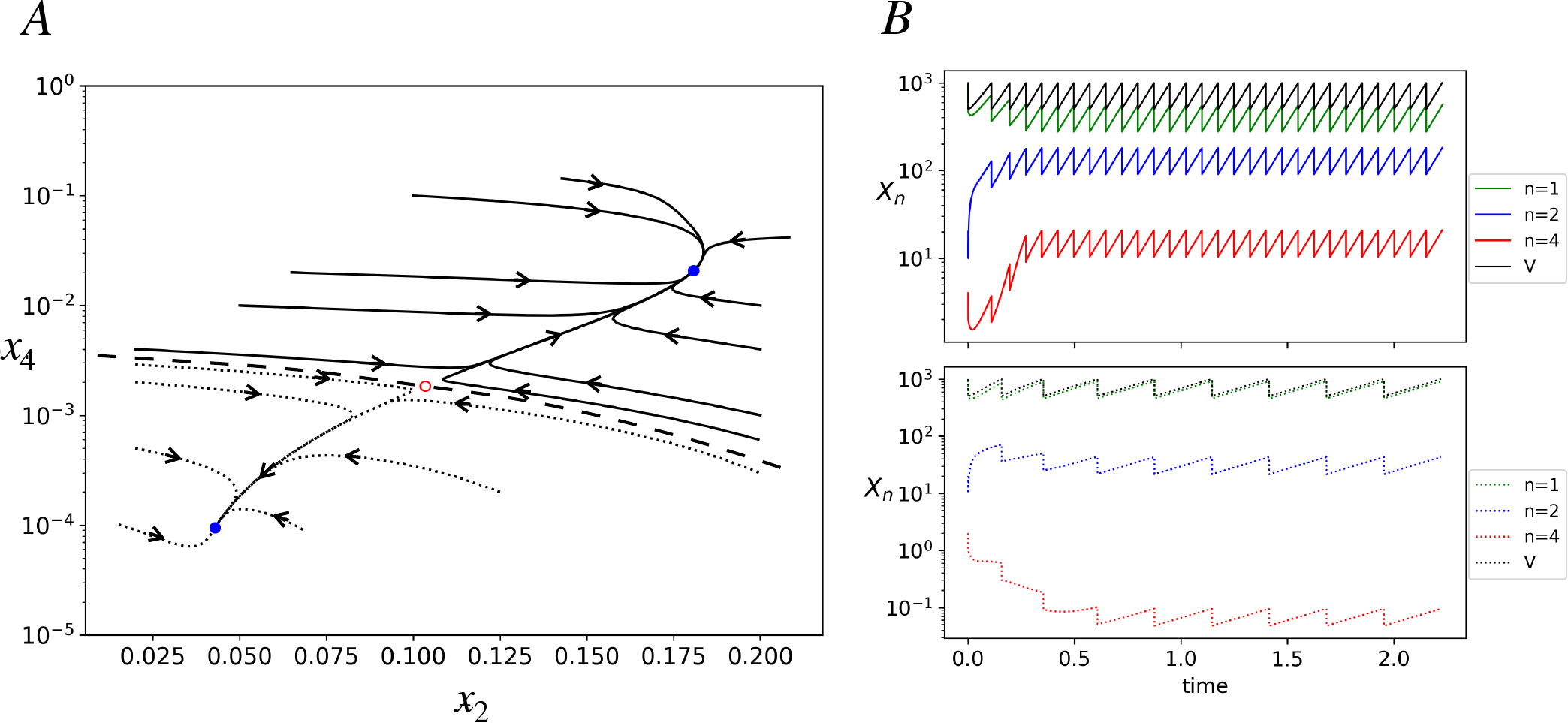}
\end{adjustwidth}
\caption{Deterministic trajectories in the bistable region of  the model. $\kappa  = 2400$, other parameters are as in Fig. \ref{fig2}. \textbf{A:} Phase portrait projected onto the $x_2-x_4$ plane. Several trajectories starting with different initial conditions are shown; they reach one of two stable fixed points denoted by blue closed dots. All the solid curve trajectories end at the stable fixed point on the top right (ACS active) while all the dotted trajectories end at the stable fixed point on bottom left of the plot (ACS inactive). The red open dot represents an unstable fixed point.  The dashed curve is a schematic of the basin boundary between the two stable fixed point attractors. \textbf{B:} Deterministic trajectories of populations (in log scale) of species $A(1)$, $A(2)$, $A(4)$ and the protocell volume as functions of time for two initial conditions. $V_c=1000$. Initial conditions: IC1 (lower panel; dotted curves): $X_1=952, X_2=20, X_4=2$. IC2 (upper panel; solid curves): $X_1=944,X_2=20,X_4=4$. Protocell starting with IC1 ends up in the \textit{inactive state} in which the population of the catalyst $A(4)$ is less than one as seen in dotted red curve in the lower panel. Protocell starting with IC2 ends up in the \textit{active state} in which the population of the catalyst is high (approximately between 10 and 20). The interdivision times in the inactive and active states are, respectively, $\tau_1 =0.269$, $\tau_2 = 0.075$.}
\label{fig3}
\end{figure}

When Eq. (\ref{eq_mu}) is substituted in Eqs. (\ref{eq_conc1}-\ref{eq_conc3}), the concentration dynamics also becomes completely defined. It has fixed points. Fig. \ref{fig2} shows a bifurcation diagram in which the fixed point concentration of $A(4)$  is plotted by varying the parameter $\kappa$. The model exhibits bistability for $\kappa^I < \kappa<  \kappa^{II}$. Note that the catalyst concentration $x_4$ in the upper stable branch is two orders of magnitude higher than in the lower stable branch. On the lower branch the rates of catalyzed reactions are smaller than the corresponding spontaneous reactions, while on the upper branch they are much higher. We therefore refer to the upper branch as one in which the ACS is \textit{active} and the lower branch as ACS \textit{inactive}. Depending on the initial condition, for a given $\kappa$ in the bistable region, the dynamics will settle into either of the two stable attractors as shown in Fig \ref{fig3}A for one such $\kappa$. For $\kappa < \kappa^I$ there is only one attractor (the inactive one), and for $\kappa > \kappa^{II}$ also only one attractor (the active one).

For each fixed point attractor, the r.h.s. of Eq. (\ref{eq_mu}) is constant. Hence in the attractor, $V$ grows exponentially, $V(t) = V(0) e^{\mu t}$ with constant $\mu$. In other words the protocell has a characteristic growth rate in each attractor given by the expression in Eq (\ref{eq_mu}). This is shown in the inset of Fig. \ref{fig2}. Hence in the bistable region, the protocell can grow with two distinct growth rates depending upon which attractor it is in. The growth rate is many times higher in the active state than in the inactive one. 

Once the concentrations have reached their fixed point attractor, (\ref{eq_mu}) implies  that $V$ grows exponentially, and Eq. (\ref{eq_vol}) then implies that each chemical population must also grow exponentially with the \textit{same} rate $\mu$. (Only if all populations grow at the same rate as $V$ will their concentrations be constant.) Thus in each attractor we have $X_i(t) = X_i(0) e^{\mu t}$. In other words, the protocell naturally exhibits \textit{balanced growth} in each attractor (growth with ratios of all populations constant \cite{campbell1957}). Exponentially growing trajectories in a nonlinear system and this remarkable emergent coordination between the chemicals without any explicit regulatory mechanism is a consequence of (a) the fact that the r.h.s. of Eqs. (\ref{eq_pop1}-\ref{eq_pop3}) are homogeneous degree one functions of the populations (if all three populations are simultaneously scaled by a factor $\beta$, $X_i \rightarrow \beta X_i$, then the r.h.s. of Eqs. (\ref{eq_pop1}-\ref{eq_pop3}) also scales by the same factor $\beta$), and (b) that the ACS structure couples all chemicals to each other. This is discussed in detail in ref. \cite{pandey2020} in the context of models of bacterial physiology.
 
Fig. \ref{fig3}B shows, for a protocell, the trajectories of its chemical populations and volume as functions of time for two very close initial conditions (defined by the population of species A(1), A(2) and A(4)) that lie in different attractor basins. They converge to different attractors: ACS-active (upper panel) and inactive (lower panel). After a protocell divides we track one of its daughters. The attractor is a fixed point for concentrations (Fig. \ref{fig3}A) but a limit cycle for populations and the volume (Fig. \ref{fig3}B). The growth phase of the limit cycle has the same constant slope for all populations in a given attractor, signifying exponential growth with the same growth rate for all chemicals in the attractor. The slope is larger (and interdivision time shorter) for the active attractor. At division, since populations and the volume both halve, concentrations do not see any discontinuity.  

The existence of bistability is robust in parameter space. It may be noted that a nonzero degradation rate $\phi$ of the dimer and tetramer is essential for bistability (as also found in the model studied in ref. \cite{giri2012}). A degradation term $\phi'x_1$ for the monomer can also be introduced in Eq. (\ref{eq_conc1}); however it is  found that $\phi'$ must be sufficiently smaller than  $\phi$ for bistability to exist.

\subsection{Stochastic dynamics of a single protocell: transitions between states of different growth rates}
We now consider the protocell under the stochastic chemical dynamics framework. The chemical populations are now non-negative integers and each unidirectional reaction occurs with a probability that depends on the populations of the reactants and the values of the rate constants. We simulate the stochastic chemical dynamics of the protocell using the Gillespie algorithm \cite{gillespie1976}. The reaction probabilities are listed in the Appendix \ref{A1}. Whenever a reaction occurs the populations of its reactants and products are updated. For large populations when fluctuations are ignored, the above mentioned probabilities lead to the deterministic Eqs. (\ref{eq_conc1}-\ref{eq_conc3}) or (\ref{eq_pop1}-\ref{eq_pop3}). In using the Gillespie algorithm for expanding volumes the rate of increase of volume needs to be taken into account \cite{lu2004,carletti2012}. In the present work since volume is treated as a function of populations (\ref{eq_vol}), we assume that it is instantaneously updated when the populations are.

\begin{figure*}
\begin{adjustwidth}{-2cm}{-2cm}
\includegraphics[width=18cm]{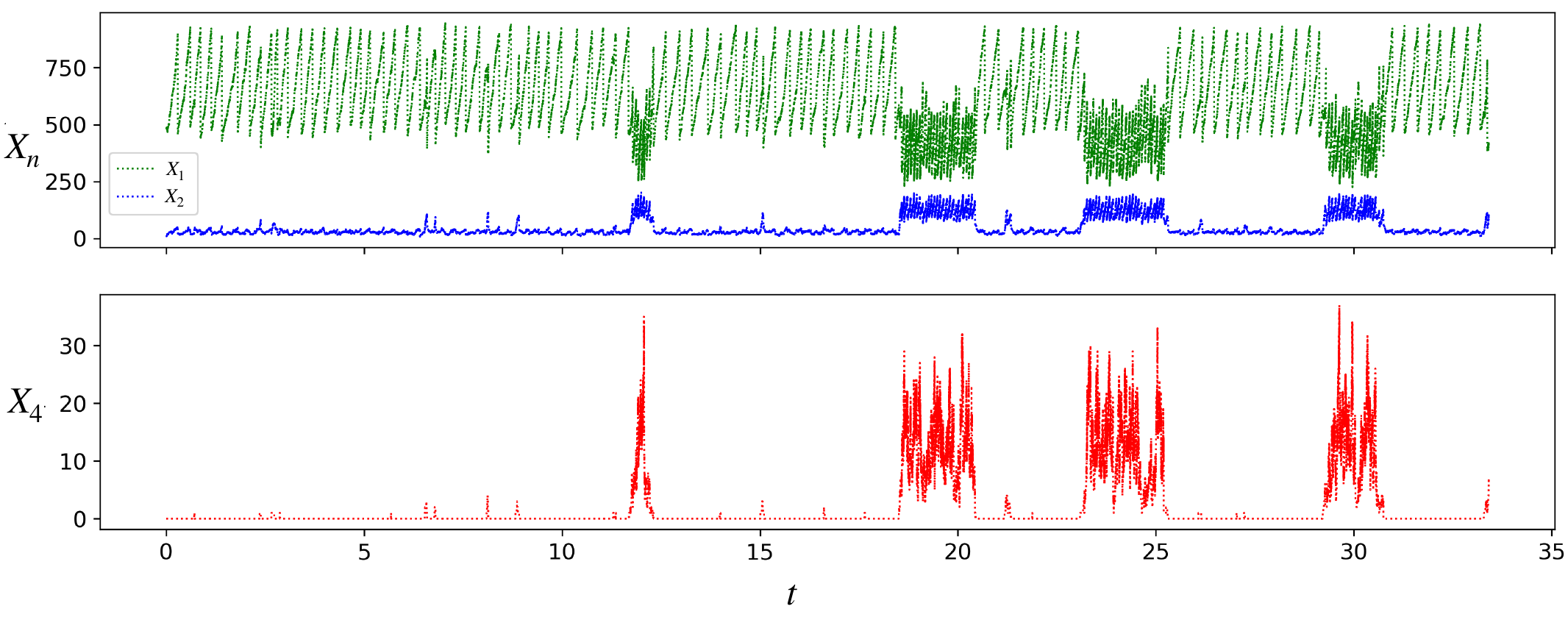}
\end{adjustwidth}
\caption{Stochastic simulation of the populations of species A(1), A(2) and A(4) for a single protocell lineage in the model. Parameter values are as in Fig. 3, $V_c=1000$. Initial condition: $X_1=480,\;X_2=10,\;X_4=0$. Note the transitions of the protocell between the inactive and active states. From a long such simulation we find that the average interdivision times in the inactive and active states are, respectively, $\langle \tau_1\rangle = 0.295$, $\langle \tau_2\rangle = 0.077$, while the average residence times in the two states are $\langle T_1\rangle = 3.413$, $\langle T_2 \rangle = 1.916$.}
\label{fig4}
\end{figure*}

Fig. \ref{fig4} shows a simulation run of the stochastic chemical dynamics of a single growing and dividing protocell. At the volume threshold $V_c$, when the protocell divides into two daughter protocells, we implement partitioning stochasticity, namely, each molecule in the mother is given equal probability of going into either daughter. In Fig. \ref{fig4}, at each division we randomly discard one of the two daughters and choose one for further tracking, in order to display a single-cell trajectory over several divisions (effectively it is the trajectory of a single lineage of protocells).

Starting from the initial condition shown where the protocell is composed of only $A(1)$ and $A(2)$, the protocell initially grows and divides in the inactive state. The first $A(4)$ molecule is produced by the chance occurrence of the uncatalyzed reaction 2. Production of a sufficient number of $A(4)$ molecules triggers a transition to the active state, where the population of $A(4)$ is significantly larger than in the inactive state. As in Fig. \ref{fig3} for the deterministic case, so also in Fig. \ref{fig4} it can be seen that the protocell in the active state grows and divides faster than in the inactive state. However, unlike the deterministic case, we also see transitions between the inactive and active states. These transitions occur because for a small protocell ($500 \leq V \leq 1000$ for the protocell in Fig. \ref{fig4}), chance production or depletion of a few molecules of $A(4)$ is enough to push its concentration into the basin of the other attractor. Note that in Fig. \ref{fig4} the protocell lineage spends more time in the inactive state than the active. The residence times of a protocell lineage in the two attractors ($T_{\text{1}}, T_{\text{2}}$) have distributions (see Fig \ref{fig-A1} in Appendix \ref{A2}) that vary with parameters. 

Note that typically a daughter naturally inherits the state of the mother protocell: since the two daughters have roughly half the number of molecules of each type as the mother, and hence also half the volume, they have the same concentration of each chemical as the mother. Partitioning stochasticity occasionally results in a daughter losing the mother's state.

\subsection{Protocell population dynamics: Dominance of the autocatalytic state}
\begin{figure}[t]
\centering
\includegraphics[width=\textwidth]{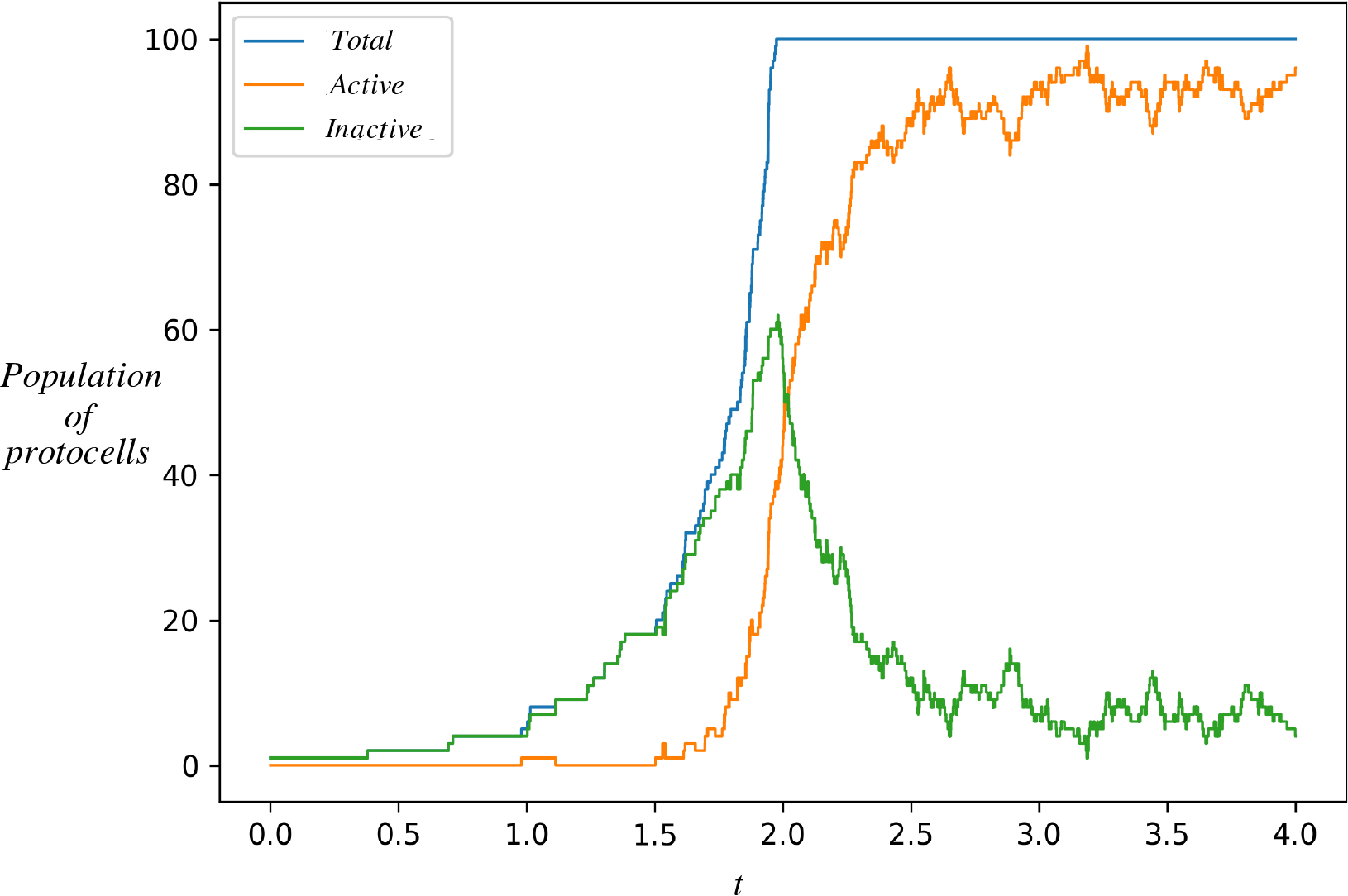}
\caption{Time evolution of a population of protocells starting from a single protocell in the inactive state. Shown is the number of protocells in the inactive state (green), active state (orange), and their sum (blue). As inactive protocells grow and divide, their population increases. The orange curve departs from zero when one of the inactive protocells makes a stochastic transition to the active state. The two populations have different growth rates. After the total population reaches an externally imposed ceiling $K$ (=100 in this figure), upon each cell division a randomly chosen protocell is removed from the population. The population eventually settles down in a stochastic steady state dominated by the active protocells. This is the natural selection of an autocatalytic state. Parameter values are as in Fig. \ref{fig4}, $K=100$.}
\label{fig5}
\end{figure}

Fig. \ref{fig5} shows the time evolution of a population of such protocells. At $t=0$ we start from a single protocell in the inactive state, whose dynamics was shown in Fig. \ref{fig4}. However, in this simulation, when a protocell divides, instead of discarding a daughter, we keep it in the simulation until the total population of protocells reaches an externally imposed ceiling $K$. After the total number of cells reaches $K$, the total population is kept constant. This done by removing one randomly chosen protocell from among the $K+1$ protocells whenever any protocell divides. Each protocell in the population is independently simulated by the single cell stochastic dynamics (Gillespie algorithm). Fig. \ref{fig5} tracks only the number of protocells in each state (active or inactive) as a function of time. 

The number of protocells in the inactive state increases whenever one of them divides. Eventually one of them makes a stochastic transition to the active state, whereupon the number of active protocells jumps from zero to one. Active protocells also make stochastic transitions to the non-active state on a certain time scale. However, since active protocells divide faster (as seen in Fig. \ref{fig4}), their number grows faster and their population catches up and overtakes the inactive population in Fig. \ref{fig5}. Eventually the active protocells dominate the population.

The curves in Fig. \ref{fig5} represent the net result of stochastic transitions and proliferation by division. The fraction of protocells in each state is expected to reach a stochastic steady state (see below) that represents a balance between proliferation and transition. In the simulations we find that the fraction of inactive cells declines when the total population hits $K$ (see Fig. \ref{fig5}). It eventually reaches its steady state fraction. A decline is seen at the time the total population hits $K$, because in this simulation at that time the fraction of inactive cells is higher than its steady state fraction (this is a consequence of the initial condition, the fact that at $t=0$ we started from a single protocell in the {\it inactive} state). In Supplementary Material Section 2 a similar qualitative behaviour can be seen for other values of $\kappa$ within the bistability region.

An approximate (mean field) model of the protocell population dynamics (valid for large populations) with no ceiling ($K \rightarrow \infty$) is the following: 
 \begin{align}
\frac{dn_1}{dt}=& \mu_1 n_1 - \lambda_1 n_1 + \lambda_2 n_2,\label{eq_mf1}\\ 
\frac{dn_2}{dt}=& \mu_2 n_2 - \lambda_2 n_2 + \lambda_1 n_1, \label{eq_mf2}
\end{align}
where $n_1(n_2)$ is the population of protocells in the inactive (active) state, $\mu_1 = \frac{\ln{2}}{\langle \tau_1 \rangle}$ and $\mu_2 = \frac{\ln{2}}{\langle \tau_2 \rangle}$ are the average growth rates of the protocell in the inactive and active states respectively, and  $\lambda_1 = \frac{1}{\langle T_1 \rangle}$ and $\lambda_2 = \frac{1}{\langle T_2 \rangle}$ are the transition rates, respectively, from the inactive to active and active to inactive states. This is a linear dynamical system $\frac{dn}{dt} = A n$, where $n=(n_1\;\;  n_2)^T$ is the column vector of protocell populations, and 
\begin{equation}
	A = 
	\begin{pmatrix}
\mu_1 - \lambda_1 & \lambda_2  \\
\lambda_1 & \mu_2 - \lambda_2  \\
\end{pmatrix}.
\end{equation}
Eqns. \mbox{(\ref{eq_mf1}-\ref{eq_mf2})} for the populations of inactive and active protocells are identical to the model used to describe the populations of persister and normal cells of bacteria \cite{balaban2004}.

The steady state fraction $f$ of active protocells in the population $f \equiv n_2/(n_1+n_2)$ can be computed from the eigenvector of $A$ corresponding to its largest eigenvalue, $e_1$. The result is:
\begin{equation}
f = \frac{\lambda_1}{e_1 + \lambda_1 + \lambda_2 - \mu_2}, \label{eq_f}
\end{equation}
where $e_1 = \frac{1}{2}[\text{tr}(A)+ \sqrt{(\text{tr}(A))^2 - 4\, \text{det}(A)}]$, $\text{tr}(A) =\mu_1 - \lambda_1 +\mu_2 - \lambda_2$ and $\text{det}(A)= (\mu_1 - \lambda_1)(\mu_2 - \lambda_2)\;-\;\lambda_1 \lambda_2$. A calculation of $f$ for a finite but large ceiling $K$ is given in the Appendix \ref{A3_1} and yields the same answer as (\ref{eq_f}), independent of $K$.

Using the averages given in the caption of Fig. 4 to determine the components of $A$, this calculation yields $f=0.925 \pm 0.014$ (mean $\pm$ standard error), with the error arising from the finite sample estimation of the averages. This agrees with the fraction found (over long times) in the stochastic steady state of the simulation of Fig. \ref{fig5}, namely $0.937 \pm 0.019$ (mean $\pm$ standard deviation). The Supplementary Material Section 3 shows the agreement between simulations and the mean field model at other values of $\kappa$. 

Note in Fig. \ref{fig5} that even though the active protocells have a higher growth rate than the inactive, a finite fraction of the inactive still survives in the steady state. This is because of the nonzero transition probability $\lambda_2$ from the active to the inactive state. If $\lambda_2$ had been zero, the eigenvector of $A$ corresponding to its largest eigenvalue would have been $(0 \;\; 1)^T$ implying that the inactive state is extinct in the steady state. When $\lambda_2 \neq 0$, one can show (see Appendix \ref{A3_2}) that if 
\begin{equation}
\lambda_2 \ll \mu_2 - \mu_1 + \lambda_1, 
\end{equation} 
then $f \simeq 1- \frac{\lambda_2}{\mu_2 - \mu_1 + \lambda_1}$ is close to unity. The quantity $1/(\mu_2 - \mu_1 + \lambda_1)$ defines a time scale of the single protocell dynamics. The above condition means that if the average lifetime $\langle T_2 \rangle$ ($=\frac{1}{\lambda_2}$) of the active state is much larger than this time scale, ACS active protocells will come to dominate the population. Another way of writing this condition is $\mu_2 \langle T_2 \rangle - \mu_1 \langle T_2 \rangle + \frac{\langle T_2 \rangle}{\langle T_1 \rangle} \gg 1$. 
Therefore a sufficient condition for active protocells to dominate is that the active protocell divides many times in its typical lifetime ($\mu_2 \langle T_2 \rangle \gg 1$) \textit{and} grows much faster than the inactive protocell ($\mu_2 \gg \mu_1$).

Note also that a nonzero $\lambda_1$ is what ensures that even if we start with a zero population of active protocells, one active protocell will sooner or later be produced by chance, leading eventually to a fraction $f$ of active protocells.

\section*{Discussion}
In this work we have constructed an example that shows (i) how autocatalytic sets of reactions inside protocells can spontaneously boost themselves into saliency and enhance the populations of their product molecules including catalysts, and (ii) how such protocells (where the ACS is active) can come to dominate in a population of protocells. Encasement within protocells serves two important functions. (i) The small size of a protocell allows a small number fluctuation of the catalyst molecules to take their concentration past the basin boundary of the attractor in which the ACS is inactive into the basin of the active attractor, thereby causing the protocell to transition from an inactive to active state. A large container would require a larger number fluctuation to achieve the same transition, which is more unlikely. (ii) Protocells in the active state grow at a faster rate than the inactive state, thereby eventually dominating in population. The differential growth rate is a consequence of the fact that the protocell size depends upon its internal chemical populations, a possibility that is precluded when we discuss chemical dynamics in a fixed size container. Therefore, in this example, protocells aid both the generation and the amplification of autocatalytic sets.

The differential growth rates of the two states are not posited exogenously, but arise endogenously within the model from the underlying chemical dynamics defined by Eqs (\ref{eq_pop1}-\ref{eq_vol}) (and their stochastic version). The additional assumption made is that upon reaching a critical size a protocell divides into two daughters that share its contents. This property can arise naturally due to some physical instability. Collectively these assumptions lead to the properties of heredity, heritable variation (the variation is heritable because once the fluctuation pushes it into a new basin of attraction a protocell typically descends into its new attractor in a short time), and differential fitness in a purely physico-chemical system. This leads to the dynamics of the two subpopulations of protocells shown in Fig \ref{fig5} which is similar to that of natural selection. (A difference is that the slower growing subpopulation never goes completely extinct, due to the non-zero probability of transition of a faster growing protocell into a slower growing one.) 

The process of going from an initial state with no ACS to its establishment in a population of protocells, discussed here, might be considered the first step in the evolution of the ACS. One might wonder how the ACS would evolve further from there. It has been shown that chemistries containing ACSs exhibit multistability in fixed sized containers. In some of these chemistries simpler ACSs involving small catalyst molecules are nested inside more complex ACSs having larger and more efficient catalyst molecules \cite{giri2012}. The multiple attractor states correspond to ACSs with progressively larger molecules and higher level of complexity being active. It is possible that by embedding such chemistries within protocells, the mechanism discussed here could allow one to realize a punctuated evolutionary path through sequentially more complex ACS attractors to a state of high chemical complexity from an initial state that only contains small molecules and no ACS. This is a task for the future.

The specific artificial chemistry and protocell properties studied here are highly idealized ones. The object was to demonstrate a mechanism in principle. However, we believe the mechanism is quite general and it should be possible to demonstrate it in other models (e.g., \cite{vasas2012,villani2014}) provided multistability in a fixed environment and the emergence of distinct timescales as discussed in the present work can be established. We remark that though we have been primarily thinking of protocells as vesicles (motivated by similar models of bacterial physiology), some of our methods might also be useful in the context of micelles. Recently Kahana et al \cite{kahana2023} presented a model of the stochastic dynamics of lipid micelles which had multiple attractors corresponding to distinct composomes. It would be interesting to compare the growth rates of micelles in different attractors as well as the transition rates between the attractors in their model. 


We note that there have been independent experimental developments in constructing bistable autocatalytic chemistries \cite{maity2019} and self-replicating protocells \cite{luisi_book}. It is also established that small peptides exhibit catalytic properties \cite{gorlero2009} and they can be encapsulated within protocells to promote protocellular growth \cite{adamala2013}. A recent paper also shows the coupling of a simple autocatalytic reaction with the compartment growth and division \cite{lu2023}. A synthesis of these approaches might result in the experimental realization of the mechanism described in the present work.  

We have considered dynamics at three levels: One is the chemical dynamics of molecules within a single protocell. This depends upon molecular parameters such as rate constants, efficiency of the catalyst molecule, etc. From this we extracted effective parameters at the second level: that of a single protocell (growth rates of the two protocell states, residence times, etc.). These were then used to derive the dynamics at the third level consisting of the population of protocells. This enabled an understanding of the conditions under which active protocells would dominate. Such an approach might be useful in other settings, for example in understanding certain aspects of bacterial ecology from molecular models of single bacterial cells.  

\section*{Acknowledgements}
This research was partially supported by the Indo French Centre for the Promotion of Advanced Research (IFCPAR) project No. 5904-3. AYS would like to thank the University Grants Commission, India for a Senior Research Fellowship and a Junior Research Fellowship. We thank Sandeep Krishna, Philippe Nghe, Parth Pratim Pandey, Shagun Nagpal Sethi, Yashika Sethi and Atiyab Zafar for fruitful discussions. We would like to acknowledge the hospitality of the International Centre for Theoretical Sciences, Bengaluru and the International Centre for Theoretical Physics, Trieste, where part of this work was done.

\begin{appendices}
\section[\appendixname~\thesection]{Reaction probabilities used in Gillespie Algorithm}\label{A1}
\begin{table}[H]
	\begin{adjustwidth}{-1cm}{-1cm}
		\newcolumntype{C}{>{\centering\arraybackslash}X}
		\newcolumntype{c}{>{\hsize=.5\hsize}X}
		\begin{tabularx}{17cm}{CcCc}
			\toprule
			Reaction &Reaction Type&Reaction Probability& Deterministic\\
			 & &per unit time&rate of reaction\\
			\midrule
			\hline 
\small{$A(1)_{ext} +A(2)\; \overset{\alpha X_2}{\underset{}{\longrightarrow}}A(1) +A(2)$}&transport& $\alpha X_2$&$\alpha X_2$\\ 
\hline
\small{$A(1) +A(1)\; \overset{k_{F}}{\underset{}{\longrightarrow}}A(2)$} &spontaneous&$k_F X_1(X_1-1) V^{-1}$&$k_F X_1^2 V^{-1}$\\
\hline
\small{$A(1) +A(1) +A(4) \overset{ \kappa k_F }{\underset{}{\longrightarrow}}A(2)+A(4)$}&catalysed&\small{$ \kappa k_FX_4V^{-1}X_1(X_1-1) V^{-1}$}&\small{$\kappa k_F X_4V^{-1}X_1^2V^{-1}$}\\
\hline
\small{$A(2)+A(2)\; \overset{k_{F}}{\underset{}{\longrightarrow}}A(4)$} &spontaneous&$k_F X_2(X_2-1) V^{-1}$&$k_F X_2^2$\\
\hline
\small{$A(2)+A(2) +A(4) \overset{ \kappa k_F }{\underset{}{\longrightarrow}}A(4)+A(4)$}&catalysed&\small{$ \kappa k_FX_4V^{-1}X_2(X_2-1) V^{-1}$}&\small{$\kappa k_F X_4V^{-1}X_2^2$}\\
\hline
\small{$A(2)\overset{k_{R}}{\underset{}{\longrightarrow}} A(1) +A(1)$}&spontaneous&$k_RX_2$&$k_RX_2$\\
\hline
\small{$A(2)+ A(4) \overset{\kappa k_{R}}{\underset{}{\longrightarrow}} A(1) +A(1) +A(4)$}&catalysed&$\kappa k_R  X_4V^{-1}X_2$&$\kappa k_R X_4V^{-1}X_2$\\
\hline
\small{$A(4)\overset{k_{R}}{\underset{}{\longrightarrow}} A(2) +A(2)$}&spontaneous&$k_RX_4$&$k_RX_4$\\
\hline
\small{$A(4)+ A(4) \overset{\kappa k_{R}}{\underset{}{\longrightarrow}} A(2) +A(2) +A(4)$}&catalysed&$ \kappa k_R X_4V^{-1}(X_4 -1)$&$\kappa k_R X_4 ^2V^{-1}$\\
\hline
\small{$A(2) \overset{\phi}{\longrightarrow} \emptyset$}& degradation & $\phi X_2$& $\phi X_2$\\
\hline
\small{$A(4) \overset{\phi}{\longrightarrow} \emptyset$}& degradation & $\phi X_4$& $\phi X_4$\\
\bottomrule
\end{tabularx}
\end{adjustwidth}
\caption{List of unidirectional reactions in the model and their reaction probabilities per unit time. $V$ appearing in the table is given by the r.h.s. of the equation (\ref{eq_mu}) in the main paper text. Detertministic rate of reactions in the last column are given as in Eqs (\ref{eq_pop1}-\ref{eq_pop3}) of the main paper.}
\label{tab1}
		
\end{table}

\section[\appendixname~\thesection]{Single cell residence time and interdivision time distributions for active/inactive states of the protocell}\label{A2}
\subsection[\appendixname~\thesubsection]{Definition of an active/inactive state of a protocell}\label{definition}

In order to obtain residence times and inter-division times in the active and inactive states of the protocell, an inference has to be made from the  intracellular populations about the current state of the cell -- whether it is active or inactive. The protocell was defined to be in the active (inactive) state if its concentration profile was in the basin of attraction of the active (inactive) attractor. The two basins are separated by a basin boundary (as shown by the black dashed curve in Fig. \ref{fig3}A of the main paper). One might choose an alternative criterion based on `closeness' to the attractor state, but for the purposes of the present work, the above definition is useful. In practice, for simplicity in the present work, the concentration of the catalyst molecule ($x_4$) at the unstable fixed point (through which the basin boundary passes) was taken to be the threshold value for determining the state of the protocell. If $x_4$ was above this value, the cell was labelled as active, otherwise it was labelled as inactive. This is an approximate implementation of the above definition. While the actual values of transition times would change when the above definition is implemented exactly, we do not expect our qualitative conclusions to depend significantly on this approximation. 

\subsection[\appendixname~\thesubsection]{Definition of residence time and interdivision time} 
While tracking the trajectory of a single lineage of cells (as shown in Fig. \ref{fig4} of the main paper) the state of the protocell (1 if active; 0 if inactive) was determined for the mother protocell at every division along with the time of the division event. (For more details on data generation, see Section 4 of the Supplementary Material.) A long such trajectory gave a sequence of division times and a corresponding sequence of ones and zeros. A contiguous subsequence consisting of only ones bordered by zeros (only zeros bordered by ones) at both ends of the subsequence was declared to be an instance of residence in the active state (inactive state). The duration of such a subsequence (equal to the difference between the ending and starting times of the subsequence as measured by the corresponding division times) was taken to be the lifetime of the state. Within an active or inactive subsequence, the difference between two consecutive division times was taken to be an instance of an  interdivision  time in that  state.

\begin{figure}[H]
\centering
\includegraphics[width=14cm]{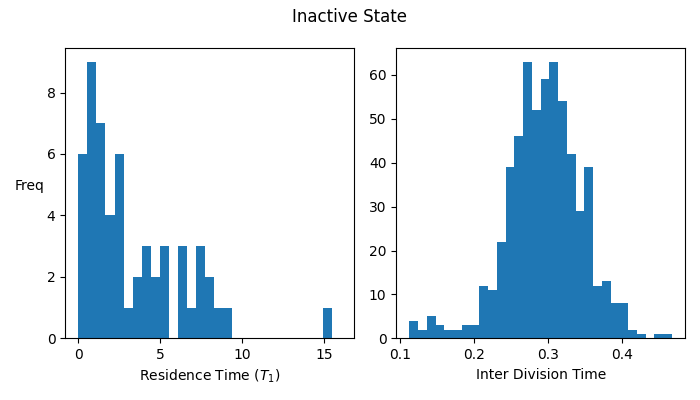}
\includegraphics[width=14cm]{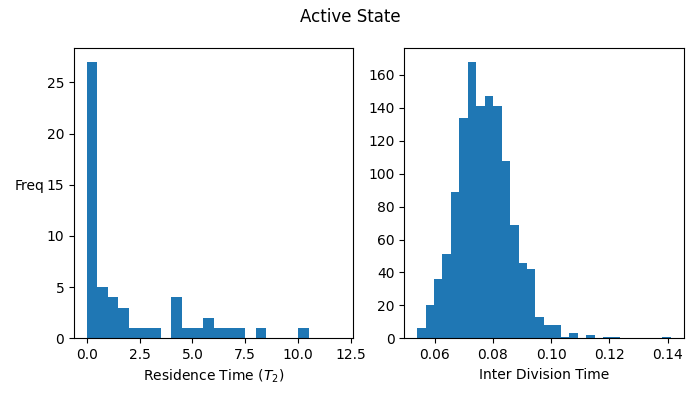}
\caption{Distribution of residence times (time spent) and inter-division time in active and inactive states. Data was collected by simulating a single lineage of growing and dividing protocells over 2000 division cycles. Parameter values: $\kappa=2400,\;k_F =1,\;\phi=20,\;\alpha=100$. The average values of the  interdivision times are $\langle \tau_1\rangle = 0.2947$, $\langle \tau_2\rangle = 0.0772$, while the average residence times in the two states are $\langle T_1\rangle = 3.413$, $\langle T_2 \rangle = 1.916$.}
\label{fig-A1}
\end{figure}

Fig \ref{fig-A1} shows the histograms for the residence times and interdivision times in the active and  inactive states using the above definitions, for one set of parameter values. 

\section[\appendixname~\thesection]{The steady state fraction of ACS Active protocells ($f$) in the protocell population}

An expression was derived for the asymptotic fraction of active protocells in the protocell population dynamics (Eq. (\ref{eq_f}) of the  main paper). The derivation used mean field equations for the populations of the active and inactive protocells and assumed indefinite growth of the two populations. Here we show that the same expression follows if we truncate the total population of protocells at a large ceiling $K$. We analyze the conditions under which this fraction is close to unity. We also present numerical  evidence that the fraction so obtained agrees with the actual stochastic simulations of protocell population dynamics at different values of $\kappa$.

\subsection[\appendixname~\thesubsection]{Calculation of $f$ for a system with finite ceiling $K$ on the total population}\label{A3_1}

In our stochastic simulations of  protocell population dynamics, the total number of protocells increases until  it reaches the ceiling $K$. After that it becomes constant because whenever a protocell divides one protocell chosen at random is removed from the population. Consider the dynamics of $n_1$ and $n_2$ (populations of the inactive and active  protocells respectively) after the total population $n_1+n_2$ has reached this constant  value $K$. If $K$ is sufficiently large, we can use the same equations as before (namely, Eqs. (\ref{eq_mf1}) and (\ref{eq_mf2}) of the main text) modified by the addition of a death term on the right hand side. In other words,
\begin{align}
\dot n_1=& \mu_1 n_1-\lambda_1 n_1 +\lambda_2 n_2-\beta n_1\\
\dot n_2=& \mu_2 n_2-\lambda_2 n_2 +\lambda_1 n_1-\beta n_2,
\end{align} 
where the last term in both equations accounts for the removal of active or inactive protocells in proportion to their existing  population (the average effect of the random removal of a protocell from the population). $\beta$ is  chosen so that the total population is constant, i.e., $n_1+n_2=K$. Then, using $\dot n_1+\dot n_2=0$, we get 
\begin{align}
\beta &=\frac{\mu_1 n_1+\mu_2n_2}{n_1+n_2} = \frac{\mu_1 n_1+\mu_2n_2}{K}.
\end{align}
Eliminating $n_1 = K-n_2$ from the $\dot n_2$  equation, and setting $\dot n_2 = 0$ to obtain a fixed point, we obtain a quadratic equation for the fixed-point value of $n_2$:
\begin{equation}
\frac{(\mu_2 -\mu_1)}{K}n_2^2-(\mu_2 -\mu_1-\lambda_2 -\lambda_1)n_2-\lambda_1 K=0. \nonumber
\end{equation}
This has the solution
\begin{align}
n_2&=\frac{K}{2(\mu_2 -\mu_1)}\left [(\mu_2 -\mu_1-\lambda_2 -\lambda_1) \pm \sqrt{(\mu_2 -\mu_1-\lambda_2 -\lambda_1)^2 +4\lambda_1 (\mu_2 -\mu_1)}\right ].\nonumber\\
\end{align}
When $\mu_2 - \mu_1$ is positive (as is the case in our simulations), the positive root must be chosen to get  a physical solution (non-negative value of $n_2$). This yields  
\begin{align}
f \equiv \frac{n_2}{K} &=\frac{1}{2(\mu_2 -\mu_1)}\left [(\mu_2 -\mu_1-\lambda_2 -\lambda_1) + \sqrt{(\mu_2 -\mu_1-\lambda_2 -\lambda_1)^2 +4\lambda_1 (\mu_2 -\mu_1)}\right ].
\end{align}
The expression of $f$ is independent of $K$. A bit of algebra shows that this expression is identical to that in Eq. (\ref{eq_f}) of the main paper. 

\subsection[\appendixname~\thesubsection]{Condition for active protocells to dominate the population}\label{A3_2}
The above expression for $f$ can be written as
\begin{align}
f &=  \frac{1}{2}\bigg[ 1 - \frac{\lambda_1 + \lambda_2}{\mu_2 - \mu_1} + \sqrt{1 + \bigg(\frac{\lambda_1 + \lambda_2}{\mu_2 - \mu_1}\bigg)^2 + \frac{2(\lambda_1 - \lambda_2)}{\mu_2 - \mu_1}}\bigg] \\
& = \frac{1}{2}[ 1 -s + \sqrt{1+s^2+2d}],
\end{align}
where $s=\frac{\lambda_1 + \lambda_2}{\mu_2 - \mu_1}$ and $d = \frac{\lambda_1 - \lambda_2}{\mu_2 - \mu_1}$.
This shows that $f$ only depends upon the two dimensionless combinations $s$ and $d$ of the four parameters.

From the above expression it immediately follows that $f=1$ when $\lambda_2 =0$, as already mentioned in the main text. We can also ask: How small should $\lambda_2$ be for $f$ to be close to unity? To see this it is useful to introduce the combinations
$z=\frac{\lambda_1}{\mu_2 - \mu_1}$ and $x=\frac{\lambda_2}{\mu_2 - \mu_1}$. Then 
\begin{equation}
f = \frac{1}{2}[1 - (z+x) + \sqrt{1 + (z+x)^2 + 2(z-x)}] = \frac{1}{2}\bigg[1 - (z+x) + (z+1)\sqrt{1 + \frac{2(z-1)x + x^2}{(z+1)^2}}\bigg].
\end{equation}
When $\frac{x}{z+1} \ll 1$, the second term inside the square root is much smaller than unity. Performing a Taylor expansion, we get $f \simeq 1-  \frac{x}{z+1}$ to leading order in $\frac{x}{z+1}$. This shows that the condition for the active protocells to dominate in the steady state of the population dynamics is 
\begin{equation}
\frac{x}{z+1} = \frac{\lambda_2}{\mu_2 - \mu_1 + \lambda_1} \ll 1.
\end{equation} 
We remark that as a special case if $x \equiv \frac{\lambda_2}{\mu_2 - \mu_1} \ll 1$, then the above condition will hold (since $\lambda_1 \geq 0$), and active protocells will dominate. However the inequality (9) gives a more general condition for active protocell domination.

\end{appendices}
\bibliography{references}

\bibliographystyle{unsrt}

\newpage
\begin{center}
 \LARGE{\bf Supplementary Material}
\end{center}
\noindent {\bf Title of paper}: Multistable protocells can aid the evolution of prebiotic autocatalytic sets \\
\noindent  {\bf Authors}:  Angad Yuvraj Singh and Sanjay Jain

\setcounter{section}{0}
\section{Robustness of the results to changes in the model structure}
In this section we present results  for a protocell model with five chemical species that relaxes certain constraints and assumptions of the model presented in the main paper in order to show the robustness of the results of the main paper. The five species include two monomers $A(1)$ and $B(1)$, two dimers $A(2)$ and $B(2)$, and one tetramer $A(4)$. Their respective populations in the protocell are denoted $X_1$, $Y_1$, $X_2$, $Y_2$ and $X_4$. The main differences are as follows:
\begin{itemize}
\item{The rate of intake of food molecules  is proportional to their difference in concentration between the outside and inside of the protocell.}
\item{There are two types  of  monomers, $A(1)$ and $B(1)$, both treated as food molecules, instead of just one.}
\item{In the model presented in the main paper, the dimer $A(2)$ was doing double duty as the enclosure forming molecule as well as a reactant to form the catalyst $A(4)$. Here the two roles  are performed by different  molecules, the enclosure  forming molecule  being the dimer $B(2)$.}
\item The definition  of the protocell volume  excludes  the population of the enclosure  forming molecule  (only  includes  populations  of molecules in the bulk of the protocell), as an  example of an alternate linear combination of chemical populations. 
\end{itemize}
While the quantitative outcomes depend upon the details, the qualitative results remain the same. These include the presence of bistability in a robust parameter region, two distinct growth rates for the two attractors, and selection of the state where the ACS is active. 

The reaction  scheme  is  as follows:
\begin{align*}
&\text{\bf Transport of}\; A(1):\;&A(1)_{ext}&\overset{\alpha Y_2}\longrightarrow A(1)\\
&\text{\bf Transport of}\; B(1):\;&B(1)_{ext}&\overset{\alpha Y_2}\longrightarrow B(1)\\
&\text{\bf R1 (uncatalyzed)}:\;&2A(1)\;&\overset{k_F}{\underset{k_R}{\rightleftharpoons}} \;\;A(2)\\
&\text{\bf R1 (catalyzed)}:\;&2A(1)+ A(4)\; &\overset{\kappa k_F}{\underset{\kappa k_R}{\rightleftharpoons}} \;\;A(2)+A(4)\\
&\text{\bf R2 (uncatalyzed)}:\;&2A(2)\;&\overset{k_F}{\underset{k_R}{\rightleftharpoons}} \;\;A(4)\\
&\text{\bf R2 (catalyzed)}:\;&2A(2)+ A(4)\; &\overset{\kappa k_F}{\underset{\kappa k_R}{\rightleftharpoons}} \;\;A(4)+A(4)\\
&\text{\bf R3 (uncatalyzed)}:\;&2B(1)\;&\overset{k_F}{\underset{k_R}{\rightleftharpoons}} \;\;B(2)\\
&\text{\bf R3 (catalyzed)}:\;&2B(1)+ A(4)\; &\overset{\kappa k_F}{\underset{\kappa k_R}{\rightleftharpoons}} \;\;B(2)+A(4)\\
&\text{\bf Degradation}:\;&A(2)\overset{\phi}\longrightarrow \emptyset,&\;\;A(4)\overset{\phi}\longrightarrow \emptyset,\;\;B(2)\overset{\phi}\longrightarrow \emptyset.
\end{align*}
In this model, the enclosure is formed by the dimers of the type $B(2)$ and is permeable only to the monomers $A(1)$ and $B(1)$. The rates at which monomers diffuse into the interior of the protocell is taken to be proportional to the number of $B(2)$ and the difference in the monomer concentrations inside and outside, $\alpha$ being the proportionality constant. The three catalyzed reactions \textbf{R1, R2, R3}, all catalyzed by $A(4)$, together with the two transport reactions, form an autocatalytic  set. The enclosure forming molecule $B(2)$ may be considered effectively a catalyst for the transport reactions. The deterministic system of equations for this model is:

\begin{align}
\frac{dX_1}{dt}= &\;\alpha Y_2 (x_1^{ext}-\frac{X_1}{V})- \:2(1+\kappa \frac{X_4}{V})(\frac{k_F X_1^2}{V}-k_RX_{2}) \\ 
\frac{dX_2}{dt}= &\;(1+\kappa\frac{X_4}{V})(\frac{k_F X_1^2}{V}-k_RX_{2})\:- \:2(1+\kappa_4 \frac{X_4}{V})(\frac{k_F X_2^2}{V}-k_RX_{4})\:-\:\phi X_2\\
\frac{dX_4}{dt}= &\;(1+\kappa \frac{X_4}{V})(\frac{k_F X_2^2}{V}-k_RX_{4}) \:-\:\phi X_4\\ 
\frac{dY_1}{dt}= &\;\alpha Y_2(y_1^{ext}-\frac{Y_1}{V}) - \:2(1+\kappa \frac{X_4}{V})(\frac{k_F Y_1^2}{V}-k_RY_{2}) \\ 
\frac{dY_2}{dt}= &\; (1+\kappa \frac{X_4}{V})(\frac{k_F Y_1^2}{V}-k_RY_{2}) -\phi Y_2,
\end{align}
where $x_1^{ext}$ and $y_1^{ext}$ are the fixed monomer concentrations outside the protocell.  The volume for this model is defined as $V=v(X_1+Y_1+ 2X_2 +4 X_4)$. This definition of $V$ excludes the population of $B(2)$. (One might imagine that $B(2)$ is a lipid molecule; once produced inside the protocell it immediately migrates to the boundary and becomes part of the enclosure, and is therefore excluded from the bulk of the protocell.) As in the main paper the protocell is assumed to divide into two equal daughters when its volume reaches the upper limit $V_c$.

We now present the behaviour of this model along the same lines as the model presented in the main paper, and show that the 5-chemical species model has the same kind of dynamics at both the single protocell and the ecosystem-of-protocells levels as the simpler model presented in the main paper. Here too there is a bistability  with the ACS active protocells having a much higher growth rate than the  ACS inactive ones; see Fig.  \ref{supp_bistable} in the Supplementary Material (SM). Under stochastic chemical dynamics the ACS can arise by chance in a single protocell that initially has no ACS (see Fig. \ref{supp_stochastic} in SM), and then the ACS active cells can take over and dominate the population of protocells (see Fig. \ref{detailed_model_f} in SM). This shows the robustness of the behaviour exhibited by the model in the main paper.

\begin{figure}[H]
\centering
\includegraphics[scale=0.05]{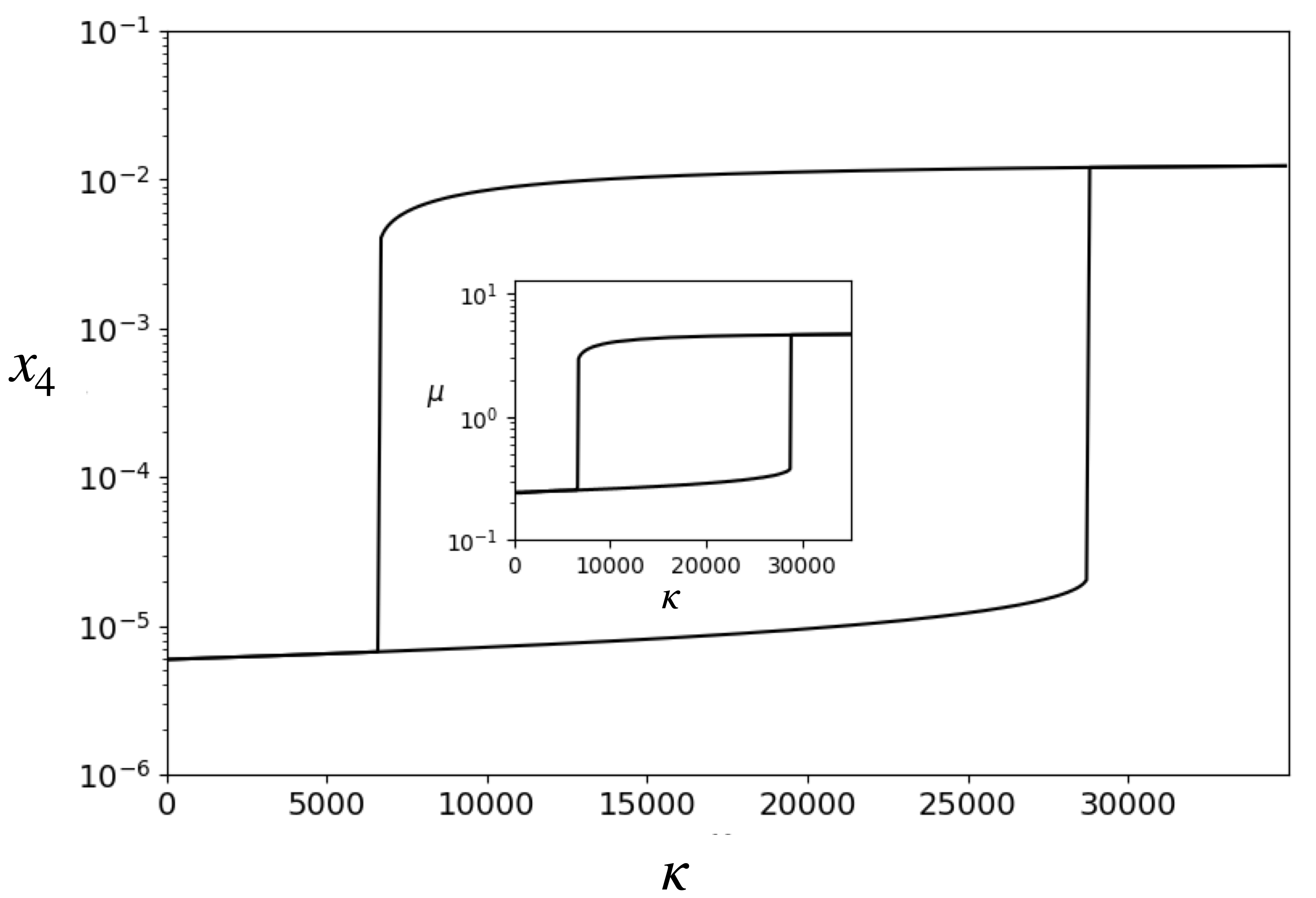}
\caption{Bifurcation diagram for the 5-chemical-species model. Parameters: $k_F=k_R=v=1,\;\phi=20,\;\alpha=100$. External concentrations: $x_1^{ext}=y_1^{ext}=1$. In the upper branch (ACS active) the catalyst has a concentration that is about three orders of magnitude higher than the lower branch (ACS inactive).  The inset shows that the growth rate of the protocell in the ACS active state is about one order of magnitude higher than in the ACS inactive state.} 
\label{supp_bistable}
\end{figure}

\begin{figure}[H]
\centering
\includegraphics[scale=0.5]{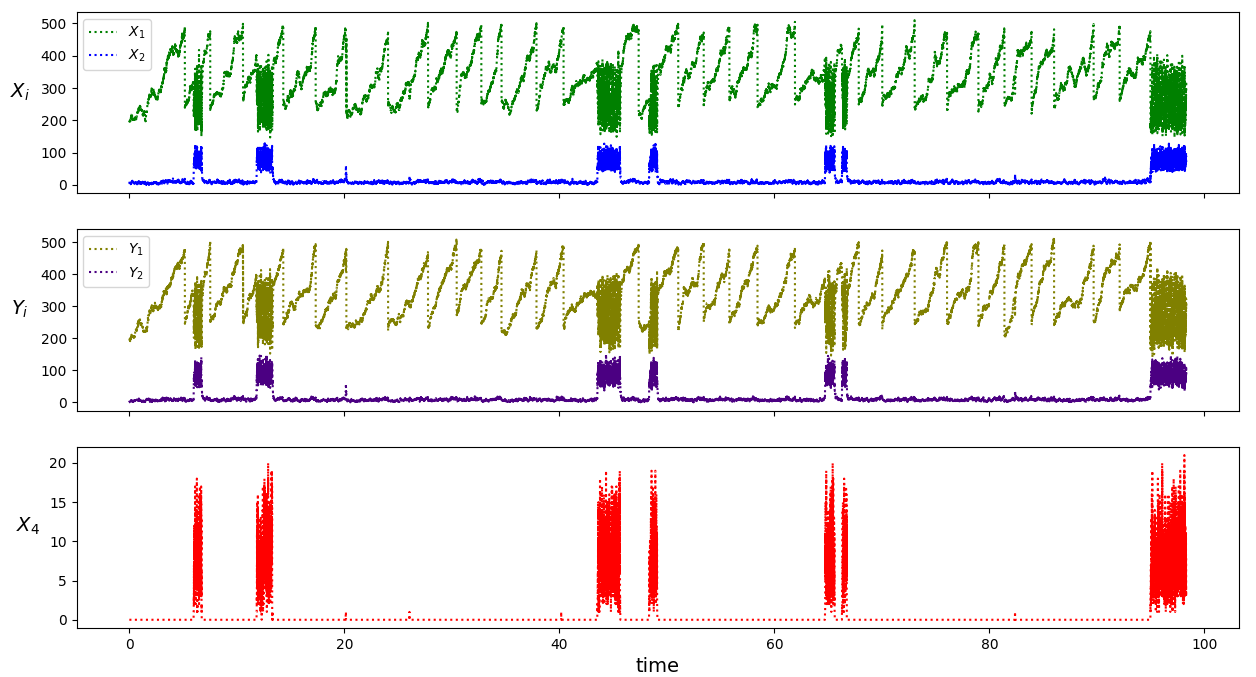}
\caption{\small Stochastic simulation of population of species A(1), B(1), A(2), B(2) and A(4) for the model using the Gillespie algorithm. Each reaction has a probability of occurrence per unit time that is related to the deterministic reaction rate along the same lines as given in Table A1 of Appendix A for the model in the main paper. Parameter values: $\kappa=15000$; rest same as in Fig. \ref{supp_bistable} of Supplementary Material. Initial condition: $X_1=Y_1=200,\;X_2=Y_2=10,\;X_4=0$. From a long such simulation we find that the average interdivision times in the inactive and active states are, respectively, $\langle \tau_1\rangle = 2.768$, $\langle \tau_2\rangle = 0.164$, while the average residence times in the two states are $\langle T_1\rangle = 20.728$, $\langle T_2 \rangle = 3.058$.}
\label{supp_stochastic}
\end{figure}

\begin{figure}[H]
\centering
\includegraphics[scale=0.8]{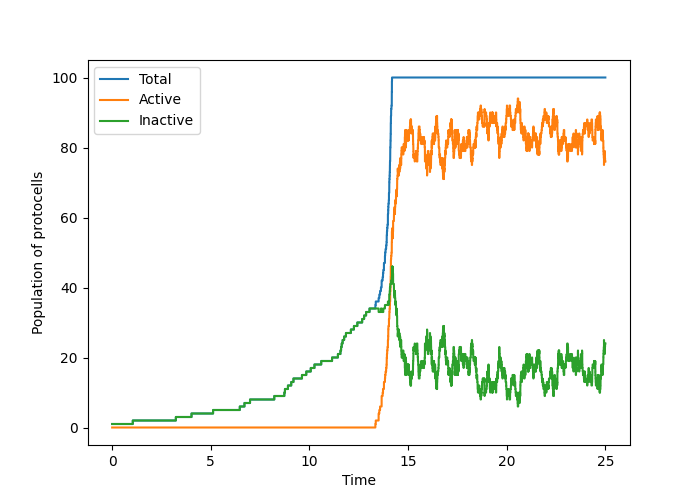}
\caption{\small Time evolution of a population of protocells in the 5-chemical-species model starting from a single protocell in the inactive state. Each individual protocell is simulated by the Gillespie algorithm for its internal chemical dynamics. Shown is the number of protocells in the inactive state (green), active state (orange), and their sum (blue). After the total population reaches an externally imposed ceiling (100 in this figure), upon each further cell division a randomly chosen protocell is removed from the population. Parameters: $\kappa=15000$, $k_F = 1$, $\phi=20$, $\alpha=100$, $V_c=1000$. Note the  domination of the active protocell population in the stochastic steady state of the protocell population dynamics, starting from an initial state with only one protocell in the inactive state.}
\label{detailed_model_f}
\end{figure}

\newpage
\section{Robustness of model behaviour at other values of catalytic efficiency ($\kappa$)}

In this section we show the model behaviour at two other values of the catalytic efficiency $\kappa$ closer to the two ends of the bistable region of $\kappa$ shown in Fig \ref{fig2} of the main paper, keeping all the other parameters the same as used to generate plots in the main paper, i.e., $k_F=1.0,\;\phi=20,\;\alpha=100,\;V_c=1000$.
\subsection{$\kappa=2000$}
\begin{figure}[H]
\centering
\includegraphics[scale=0.12]{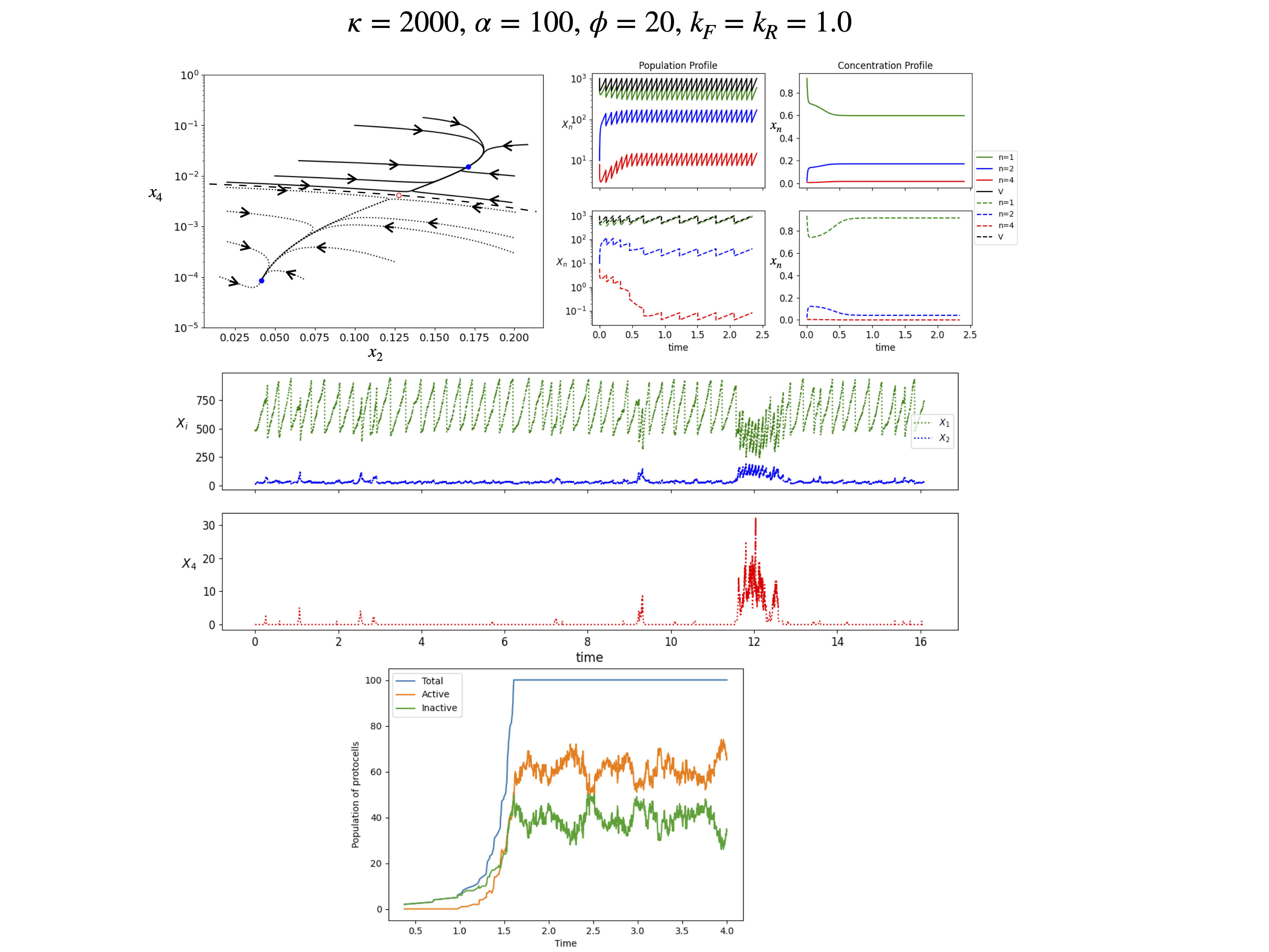}
\caption{Simulations of the model at $\kappa = 2000$. The same panels as in Figs. \ref{fig3}, \ref{fig4} and \ref{fig5} of the main paper are shown, but at $\kappa=2000$. For the deterministic case, concentrations of the molecules as a function of time are also shown.}
\label{supp_panel2000}
\end{figure}

\newpage
\subsection{$\kappa=3400$}
\begin{figure}[H]
\centering
\includegraphics[scale=0.12]{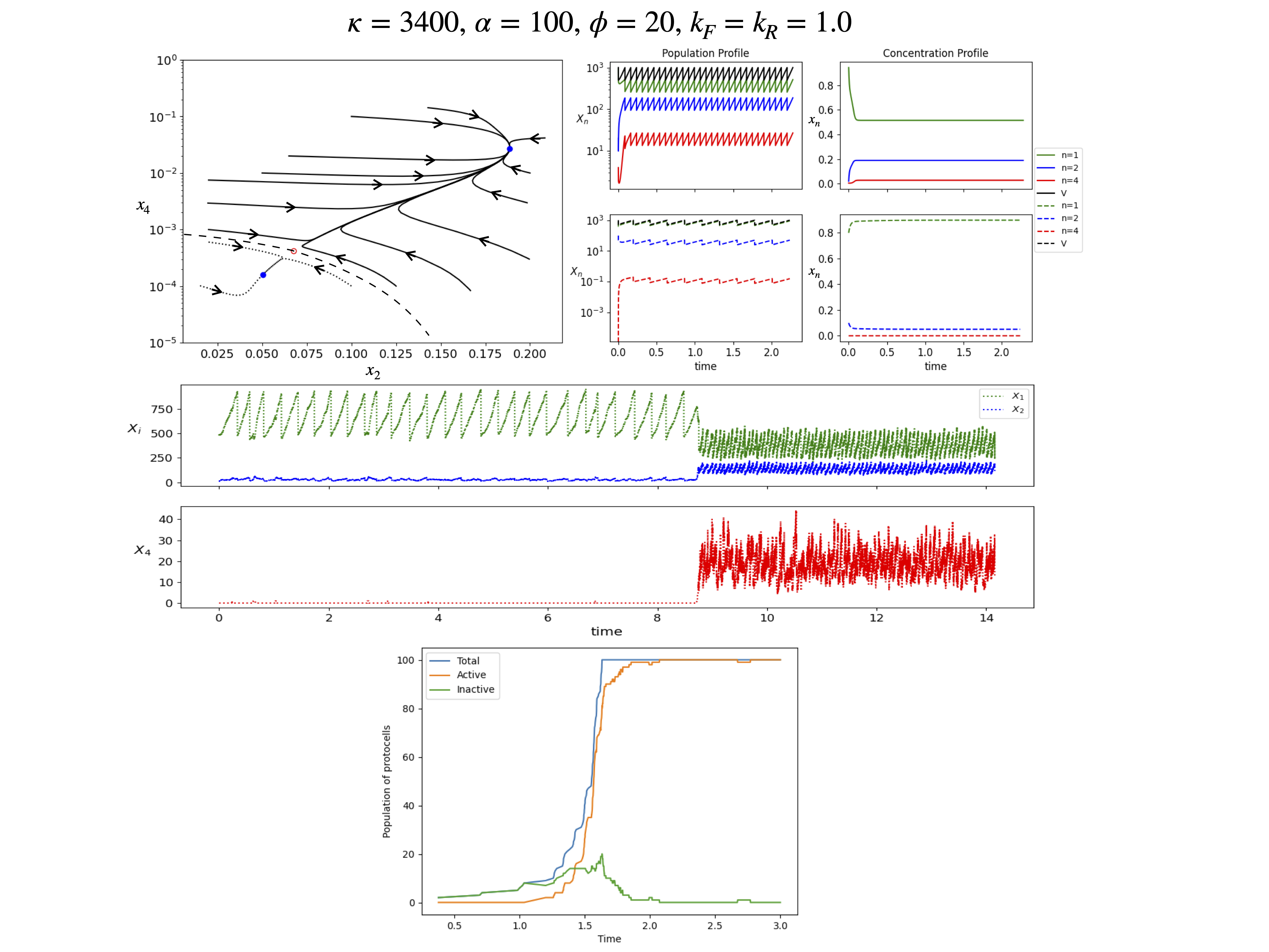}
\caption{Simulations of the model at $\kappa = 3400$. The same panels as in Figs. \ref{fig3}, \ref{fig4} and \ref{fig5} of the main paper are shown, but at $\kappa=3400$. For the deterministic case, concentrations of the molecules as a function of time are also shown.}
\label{supp_panel3400}
\end{figure}

As $\kappa$ increases within the bistable  region, the lifetime of the inactive state decreases and that of the active state increases. This  is  expected since the basin size of  the inactive attractor declines and that  of the active attractor grows as $\kappa$ increases from $\kappa^I$ to $\kappa^{II}$ (see, e.g., the difference between the unstable branch and the two stable  branches in Fig. \ref{fig2} of the main paper). This increases the steady state fraction of the active protocells in the dynamics of protocell populations. However the qualitative behaviour of the model is unchanged.

\newpage

\section{Comparison of $f$ from mean field model with $f$ obtained by simulations}
Table \ref{table-data} and Fig. \ref{fig-mf} of  the supplementary material  compare the value of $f$ obtained from the analytic expression given in Eq. (13) of main paper and in Appendix C with the value in stochastic simulations of the protocell dynamics discussed in the main paper (denoted  $f_{sim}$), at different values of the catalytic efficiency $\kappa$. 

\begin{table}
\small{
\begin{tabular}{|c|c|c|c|c|c|c|c|c|}
\hline
$\kappa$ &$\mu_1 \pm \Delta \mu_1$ & $\mu_2 \pm \Delta \mu_2$ &$\lambda_1 \pm \Delta \lambda_1$ & $\lambda_2 \pm \Delta \lambda_2$ & $f \pm \Delta f$ & $f_{sim}\pm \Delta f_{sim}$ \\
\hline
1900&2.371 $\pm$ 0.013 &8.508 $\pm$ 0.684& 0.072 $\pm$ 0.011 &3.122 $\pm$ 0.423 &0.503 $\pm$0.122&0.475 $\pm$ 0.05 \\
\hline 
2000&2.372 $\pm$ 0.013&8.752 $\pm$ 0.075 &  0.096 $\pm$ 0.015&2.205 $\pm$ 0.351&0.662 $\pm$ 0.059& 0.664 $\pm$ 0.027 \\
\hline
2200&2.324 $\pm$0.013&8.861 $\pm$ 0.045 &  0.179 $\pm$ 0.025&1.074 $\pm$ 0.152&0.841 $\pm$ 0.024 &0.854 $\pm$ 0.023 \\
\hline
2400&2.352 $\pm$ 0.017 & 8.976 $\pm$ 0.031 & 0.293 $\pm$ 0.035 &0.522 $\pm$ 0.092&0.925 $\pm $ 0.014&0.937 $\pm$ 0.019\\
\hline 
2600&2.290 $\pm$ 0.020 & 9.081 $\pm$ 0.028 & 0.300 $\pm$ 0.042 &0.314 $\pm$ 0.086&0.956 $\pm $ 0.013&0.974 $\pm$ 0.012\\
\hline 
2800&2.277 $\pm$ 0.034 & 9.112 $\pm$ 0.027 &0.359 $\pm$ 0.090 &0.144 $\pm$ 0.030&0.980 $\pm$ 0.005 &0.992 $\pm$ 0.006\\
\hline
\end{tabular}
}
\caption{Comparison of $f$ from mean field model and stochastic simulations of protocell population dynamics ($f_{sim}$) for different values of $\kappa$. Parameters: $v=k_R=1$, $k_F=1$, $\phi=20$, $\alpha=100$, $V_c=1000$. For calculating $f$ from the mean field model, the parameters $\mu_1$, $\mu_2$, $\lambda_1$, and $\lambda_2$ are estimated as discussed in the main paper as well as in Appendix B of the main paper. $f_{sim}$ data was generated from stochastic simulations of protocell dynamics in which the ceiling of the total number of protocells was taken to be $K=250$.}
\label{table-data}
\end{table}

\begin{figure}[H]
\centering
\includegraphics[width=13cm]{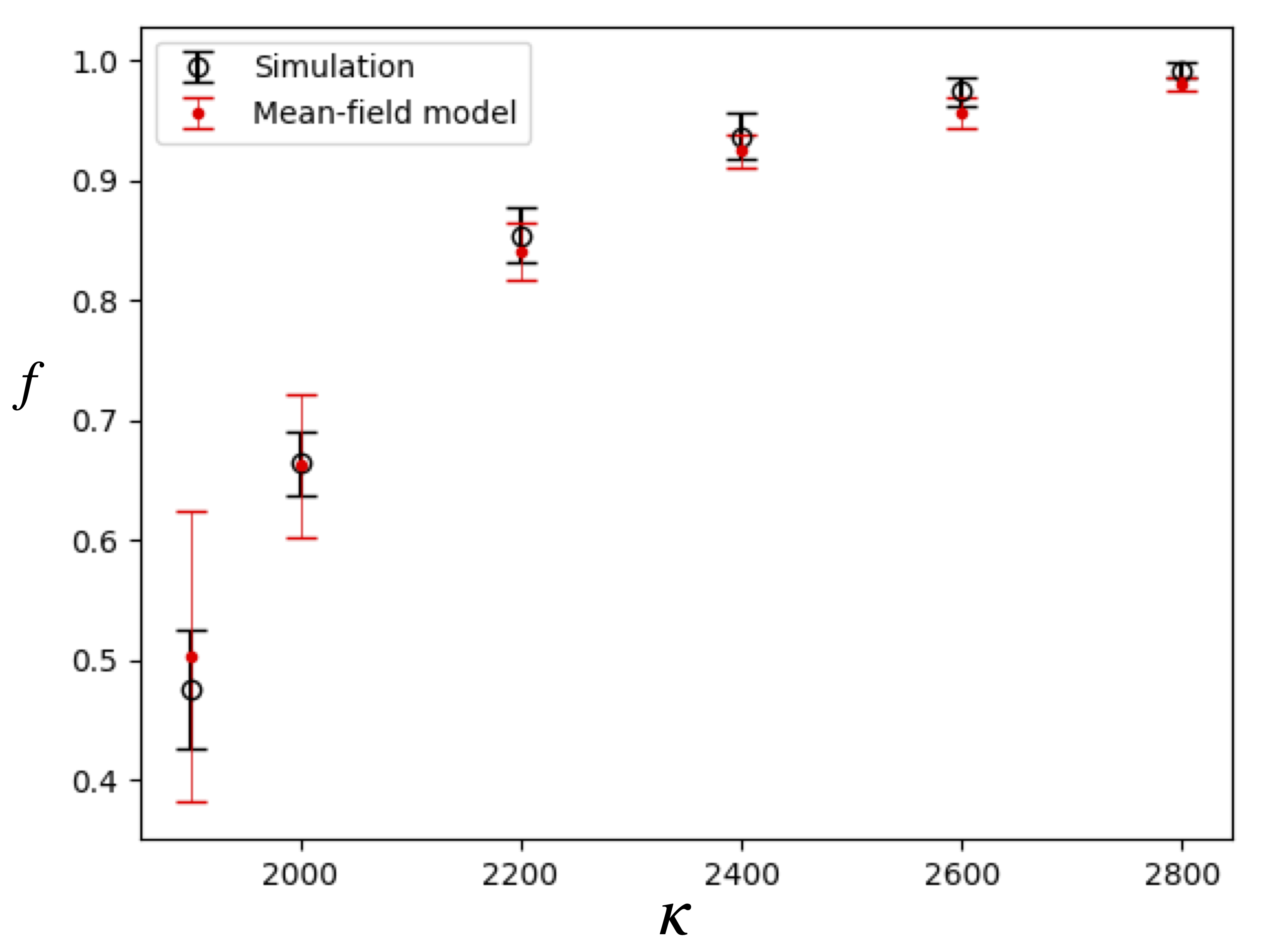}
\caption{Fraction $f$ of ‘ACS active’ protocells in the stochastic steady state of the protocell population dynamics versus $\kappa$, obtained from simulation (black hollow circles) and the mean field model (red solid dots). The error bars in $f$ and $f_{sim}$ are $\pm \Delta f$ and $\pm \Delta f_{sim}$ respectively, whose calculation is discussed in Section 3 of the Supplementary Material. Data taken from Table \ref{table-data} of Supplementary Material. $k_F = 1, \;\phi = 20, \; \alpha = 100,\; V_{c} = 1000$.}
\label{fig-mf}
\end{figure} 

The analytic value of $f$ obtained from the mean field model agrees with  $f_{sim}$ within error bars. Note that the individual parameters $\mu_1$, $\mu_2$, $\lambda_1$, and $\lambda_2$ in the analytic expression for $f$ are obtained from average values of $\tau_1$, $\tau_2$, $T_1$ and $T_2$ calculated from the respective histograms (such as those displayed in Fig. A1 of Appendix B in the main paper) generated from the stochastic simulation of a single growing and dividing protocell. Therefore, all of the parameters have errors (given in Table \ref{table-data}) arising from the standard errors of the means. The error $\Delta f$ in $f$ is computed from the analytical expression of $f$ using the above mentioned standard errors in each of the four quantities. The error $\Delta f_{sim}$ in $f_{sim}$ is just the standard deviation of $f_{sim}$ in the stochastic steady state.


\section{Data structure and cleaning methodology}
Data analysis was primarily performed using data generated from two stochastic simulations: \textbf{Stochastic single cell growth-division} and \textbf{Population of protocells}. 
\subsection{Stochastic single cell growth-division}
Following is the format of data prepared for analysing various aspects of the model:
\begin{enumerate}
\item{\textbf{Raw Data Level 0:} Raw data is first stored in the structure given in Table \ref{rd_level-0}. The raw data consists of the copy number of species ($X_i$), time at which the reaction occurred, volume of the protocell, and the generation at which the cell is when the internal reactions are happening. The generation count\footnote{Generation count is defined as the number of divisions the cell has undergone during the course of the simulation.} is set to 0 at the start of the simulation.\\
\begin{table}[H]
\centering
\begin{tabular}{|c|c|c|c|c|c|}
\hline
Generation count&Time of reaction &$V$&$X_1$&$X_2$&$X_4$\\
\hline 
\end{tabular}
\caption{Representation of data generated for Raw Data Level 0.}
\label{rd_level-0}
\end{table}
}

\item{\textbf{Raw Data Level 1:} From the Raw Data Level 0, data \textbf{ONLY} at the time of division is extracted and stored separately in the format given in Table \ref{rd_level-1}. This data has the values of species copy number and the state (in terms of binary string `0' for inactive or `1' for active) of the mother protocell at the time of division. \\
\begin{table}[H]
\centering
\begin{tabular}{|c|c|c|c|c|c|c|}
\hline
Generation count&State (0/1)&Time at division &$V$&$X_1$&$X_2$&$X_4$\\
\hline 
\end{tabular}
\caption{Representation of data generated for Raw Data Level 1.}
\label{rd_level-1}
\end{table}
Note that the time is recorded only at the point when the cell divides. This time marks the end of previous generation or start of the next generation. A transition can occur at any point within a single cell cycle. However, the state of the cell (0 or 1) is noted only at the time of division in the data above. The state of the cell is decided as per the criteria defined in Appendix B in the paper. Transient from this data is removed as per the guidlines given next. 
}

\item{\textbf{Removing the transient}: The transient trajectory of the cell is defined as the initial phase where the concentrations of the species inside the cell have not reached a stochastic steady state. To generate the data set to extract the parameters of the mean field model (residence times and interdivision times) for the two states from a run, the transient in the beginning of the run needs to be removed from data stored in Raw Data Level 1. It is typically observed that the stochastic steady state is reached within the first few division cycles. After 15 division cycles we ask: Has a transition occurred yet? There could be two possibilities. 
\begin{enumerate}
\item{\textbf{No transition has taken place in the first 15 division cycles:} Then the point where the first transition occurs is the start of data recording. This point is the first instance where the cell has changed its state (from 0 to 1 or 1 to 0). }
\item{ \textbf{A transition has taken place in the first 15 division cycles:} In this case, the first transition is ignored. Data collection starts from the second transition point irrespective of whether it is within the first 15 division cycles or not.}
\end{enumerate}
 }

\item{\textbf{Truncating data collection}: Data collection stops at the last transition point in the run. This point is the last instance where the cell changes its state (from 0 to 1 or 1 to 0). }
\end{enumerate}

The Level 1 data modified by the removal of the initial transient and truncation of the end of the run is stored separately in same format as given in Table \ref{rd_level-1}, and used to construct the histograms of single protocell parameters $\tau_1, \tau_2, T_1, T_2$.

\subsection{Population of protocells}
In this simulation, data is generated in two formats:

\begin{enumerate}
\item{\textbf{Data of individual protocell:} Each time a protocell divides, a new file is generated storing the data of one of the daughter protocells starting with the population of chemical species at birth in the format shown in Table \ref{format-pop} while the data of the other daughter cell is appended to the mother cell file. Each time a reaction occurs in any cell, the revised species count ($X_i$) is appended to the file corresponding to that cell. The total number of files is equal to the number of division cycles plus number of cells the simulation began with \footnote{For results shown in the paper, the simulation started with a single cell but one can also run the simulation starting with $n$ cells}. 

\begin{table}
\centering
\begin{tabular}{|c|c|c|c|c|c|}
\hline
State (0/1)&Time of reaction &$V$&$X_1$&$X_2$&$X_4$\\
\hline 
\end{tabular}
\caption{Representation of data generated for a particular cell in the protocell population simulation.}
\label{format-pop}
\end{table}
}

\item{\textbf{Summary Data}: A summary file is created that stores the number of active/inactive protocells at every division by extracting data of all the existing protocells at the division time points. This information is stored in the format given in Table \ref{div-summary}. This data is used to generate Fig. 5 of the main paper. 

\begin{table}[H]
\centering
\begin{tabular}{|c|c|c|c|}
\hline
Total no. of cells& Division Time&No. of Active cells& No. of Inactive Cells\\
\hline 
\end{tabular}
\caption{Format of the data stored in file containing the number of active/inactive protocells in the population.}
\label{div-summary}
\end{table}
}

\end{enumerate}

\end{document}